\documentstyle[12pt,epsfig]{article}
\newcommand{\be}{\begin{equation}}
\newcommand{\ee}{\end{equation}}
\newcommand{\beq}{\begin{equation}}
\newcommand{\eeq}{\end{equation}}
\newcommand{\ba}{\begin{eqnarray}}
\newcommand{\ea}{\end{eqnarray}}
\newcommand{\bea}{\begin{eqnarray}}
\newcommand{\eea}{\end{eqnarray}}

\begin{document}
\baselineskip=15.5pt \pagestyle{plain} \setcounter{page}{1}


\def\del{{\partial}}
\def\vev#1{\left\langle #1 \right\rangle}
\def\cn{{\cal N}}
\def\co{{\cal O}}
\newfont{\Bbb}{msbm10 scaled 1200}     
\newcommand{\mathbb}[1]{\mbox{\Bbb #1}}
\def\IC{{\mathbb C}}
\def\IR{{\mathbb R}}
\def\IZ{{\mathbb Z}}
\def\RP{{\bf RP}}
\def\CP{{\bf CP}}
\def\Poincare{{Poincar\'e }}
\def\tr{{\rm tr}}
\def\tp{{\tilde \Phi}}

\def\TL{\hfil$\displaystyle{##}$}
\def\TR{$\displaystyle{{}##}$\hfil}
\def\TC{\hfil$\displaystyle{##}$\hfil}
\def\TT{\hbox{##}}
\def\HLINE{\noalign{\vskip1\jot}\hline\noalign{\vskip1\jot}} 
\def\seqalign#1#2{\vcenter{\openup1\jot
  \halign{\strut #1\cr #2 \cr}}}
\def\lbldef#1#2{\expandafter\gdef\csname #1\endcsname {#2}}
\def\eqn#1#2{\lbldef{#1}{(\ref{#1})}%
\begin{equation} #2 \label{#1} \end{equation}}
\def\eqalign#1{\vcenter{\openup1\jot
    \halign{\strut\span\TL & \span\TR\cr #1 \cr
   }}}
\def\eno#1{(\ref{#1})}
\def\href#1#2{#2}
\def\half{{1 \over 2}}

\def\ads{{\it AdS}}
\def\adsp{{\it AdS}$_{p+2}$}
\def\cft{{\it CFT}}

\newcommand{\ber}{\begin{eqnarray}}
\newcommand{\eer}{\end{eqnarray}}

\newcommand{\beqar}{\begin{eqnarray}}
\newcommand{\cN}{{\cal N}}
\newcommand{\cO}{{\cal O}}
\newcommand{\cA}{{\cal A}}
\newcommand{\cT}{{\cal T}}
\newcommand{\cF}{{\cal F}}
\newcommand{\cC}{{\cal C}}
\newcommand{\cR}{{\cal R}}
\newcommand{\cW}{{\cal W}}
\newcommand{\eeqar}{\end{eqnarray}}
\newcommand{\th}{\theta}
\newcommand{\lm}{\lambda}\newcommand{\Lm}{\Lambda}
\newcommand{\eps}{\epsilon}
\newcommand{\pa}{\paragraph}
\newcommand{\pt}{\partial}
\newcommand{\de}{\delta}
\newcommand{\De}{\Delta}
\newcommand{\lb}{\label}


\newcommand{\nonu}{\nonumber}
\newcommand{\oh}{\displaystyle{\frac{1}{2}}}
\newcommand{\dsl}
  {\kern.06em\hbox{\raise.15ex\hbox{$/$}\kern-.56em\hbox{$\partial$}}}
\newcommand{\id}{i\!\!\not\!\partial}
\newcommand{\as}{\not\!\! A}
\newcommand{\ps}{\not\! p}
\newcommand{\ks}{\not\! k}
\newcommand{\D}{{\cal{D}}}
\newcommand{\dv}{d^2x}
\newcommand{\Z}{{\cal Z}}
\newcommand{\N}{{\cal N}}
\newcommand{\Dsl}{\not\!\! D}
\newcommand{\Bsl}{\not\!\! B}
\newcommand{\Psl}{\not\!\! P}
\newcommand{\eeqarr}{\end{eqnarray}}
\newcommand{\ZZ}{{\rm \kern 0.275em Z \kern -0.92em Z}\;}

\begin{titlepage}

\leftline{OUTP-03-24-P}
\leftline{\tt hep-th/0309161}

\vskip -.8cm


\begin{center}

\vskip 1.5 cm

{\LARGE Spinning and rotating strings for ${\cal {N}}=1$ SYM theory and brane
constructions} \vskip .3cm


\vskip 1.cm

{\large Martin Schvellinger{\footnote{
martin@thphys.ox.ac.uk,
martin@fisica.unlp.edu.ar}}}

\vskip 0.6cm

{\it Theoretical Physics, Department of Physics, University of
Oxford. \\ 1 Keble Road, Oxford, OX1 3NP, UK.}


\vskip 0.1cm

{\it Instituto de F\'{\i}sica de La Plata, CONICET and
Departamento de F\'{\i}sica, \\ Facultad de Ciencias Exactas,
Universidad Nacional de La Plata. \\
C.C. 67 (1900) La Plata,
Argentina.}


\vspace{1.7cm}

{\bf Abstract}

\end{center}

We obtain spinning and rotating closed string solutions in $AdS_5
\times T^{1,1}$ background, and show how these solutions can be mapped
onto rotating closed strings embedded in configurations of
intersecting branes in type IIA string theory. Then, we discuss
spinning closed string solutions in the UV limit of the
Klebanov-Tseytlin background, and also properties of classical
solutions in the related intersecting brane constructions in the UV
limit. We comment on extensions of this analysis to the
deformed conifold background, and in the corresponding intersecting
brane construction, as well as its relation to the deep IR limit of
the Klebanov-Strassler solution. We briefly discuss on the relation
between type IIA brane constructions and their related M-theory
descriptions, and how solitonic solutions are related in both
descriptions.

\noindent

\end{titlepage}

\newpage

\tableofcontents


\vfill

\section{Introduction}

The large $N$ limit of SYM theories is related to supergravity
solutions through AdS/CFT duality \cite{Maldacena:1997re}.
Particularly, the first and best understood example of this phenomenon
\cite{D'Hoker:2002aw} is the duality between the large $N$ limit of
${\cal{N}}=4$ $SU(N)$ SYM theory in 4d and type IIB supergravity on
$AdS_5 \times S^5$ with $N$ units of $F_5$ flux through $S^5$,
being this background the near horizon limit of a configuration of $N$
parallel D3-branes. Essentially, the system is considered in the
decoupling limit, {\it i.e.} modes propagating on the brane
world-volume and modes propagating in the bulk are decoupled.

More recent studies have shown that it is possible to extend these
ideas beyond the supergravity approximation. In particular,
certain ${\cal{N}}=4$ $SU(N)$ SYM theory operators with large
$R$-symmetry charge have been proposed to be dual to certain
closed string theory states \cite{Berenstein:2002jq}, which are
obtained by quantization of type IIB string theory on a pp-wave
background \cite{Metsaev}. The same conclusion has been reached by
quantizing string theory around certain classical solutions,
leading to semi-classical limits where the string/gauge theory
duality can be extended \cite{Gubser:2002tv}. In particular, short
spinning closed strings in $AdS_5$ have been shown to reproduce
Regge behavior.  On the other hand, long spinning closed string
solutions in $AdS_5$ reproduce the logarithmic anomalous scaling
dimension for twist two SYM theory operators
\cite{Gubser:2002tv,Frolov:2002av,
Tseytlin:2002ny,Russo:2002sr,kruczenski}. The last is a very
important feature of the UV behavior of asymptotically free
Yang-Mills theories both, supersymmetric and non-supersymmetric,
and it is believed to be an universal property
\cite{Gross:ju,Tseytlin:2002ny}.

The finding of new closed string solutions
\cite{Frolov:2002av,Tseytlin:2002ny,Russo:2002sr,kruczenski,
Iengo:2002tf,Frolov:2003qc,tseytlin,Armoni:2002xp,
Minahan:2002rc,Rashkov:2002zt,alexandrova,Ouyang:2002vg,
Buchel:2002yq,Pons:2003ci,Alihahisha:2002fi,Mateos:2003de} has
motivated to explore different regions of the spectrum by using both
sides, {\it i.e.}  weakly coupled field theory and semi-classical
string solutions. In addition, rotating closed membranes have been
recently studied
\cite{Ouyang:2002vg,Alishahiha:2002sy,Hartnoll:2002th,ryang},
obtaining generalizations of the energy-spin relations for spinning
closed strings, including the logarithmic behaviour and energy-charge
relations as well.

The interest in studying non-conformal SYM theories with less
supercharges is obvious. In fact, during the last few years, a very
extensive work has been done in order to develop the supergravity
duals of field theories in the large $N$ limit. For instance, the
large $N$ limit of ${\cal{N}}=2$ $SU(N) \times SU(N)$ SYM theory in 4d
has been proposed to be dual to type IIB supergravity on $AdS_5 \times
S^5/Z_2$ \cite{Kahru}.  From the viewpoint of ${\cal{N}}=1$
supersymmetry, the ${\cal{N}}=2$ theory contains two pairs of chiral
bifundamentals and two adjoints, coupled by a cubic superpotential.
After adding a mass perturbation and integrating out the adjoints, the
superpotential of ${\cal{N}}=1$ $SU(N) \times SU(N)$ SYM theory in 4d
is obtained. In this resulting ${\cal{N}}=1$ theory there are chiral
superfields $A_k$, $k=1, 2$ transforming in the
$(\mathrm{N},\mathrm{\overline{N}})$ representation and $B_l$, $l=1,
2$ transforming in the $(\mathrm{\overline{N}},\mathrm{N})$
representation, plus a quartic superpotential. This theory turns out
to be dual to type IIB supergravity on $AdS_5 \times T^{1,1}$ and $N$
units of $F_5$ flux through $T^{1,1}$ \cite{Klebanov:1998hh}.
D3-branes wrapped over the 3-cycles of $T^{1,1}$ can be identified
with baryon-like chiral operators built out of products of $N$ chiral
superfields, while a D5-brane wrapped over a 2-cycle of $T^{1,1}$ acts
as a domain wall in $AdS_5$ (usually called fractional D3-brane)
\cite{Gubser:1998fp}. Moreover, the dual description of $N$ regular
plus $M$ fractional D3-branes at the conifold singularity is a
non-conformal ${\cal N}=1$ $SU(N+M)\times SU(N)$ SYM theory. This
supergravity dual description reproduces the logarithmic flow of
couplings found in the field theory \cite{Klebanov:1999rd}.  In
reference \cite{Klebanov:1999rd} the leading order $M/N$ effects were
considered. Furthermore, the back-reaction of the fractional branes on
the gravitational background was studied in \cite{Klebanov:2000nc},
obtaining the logarithmic RG flow of couplings found in the field
theory at all scales. In addition, Klebanov and Strassler showed how
the gauge theory undergoes repeated Seiberg-duality transformations,
leading to a reduction of the rank of both gauge groups in $M$ units
every time. They also noted that the gauge theory confines in the IR
and, its chiral symmetry breaking removes the singularity of the
Klebanov-Tseytlin solution by deforming the conifold
\cite{Klebanov:2000hb}. More recent investigations include massive
flavored fundamental quarks in the supergravity dual of ${\cal N}=1$
SYM theory by introducing D7-brane probes to the Klebanov-Strassler
solution, finding a discrete spectra exhibiting a mass gap of the
order of the glueball mass \cite{Sakai:2003wu}\footnote{It is also very
interesting the computation of the meson spectrum of an ${\cal {N}}=2$
SYM theory with fundamental matter from its dual string theory on
$AdS_5 \times S^5$ with a D7 brane probe done in
ref.\cite{kruczenski2}.}.  On the other hand, inspired in
\cite{Gubser:2002tv}, semi-classical solutions of rotating closed
strings in both, $AdS_5$ and $T^{1,1}$ spaces have been studied
\cite{Tseytlin:2002ny}. In particular, in ref.\cite{Buchel:2002yq}
classical string solutions were considered in the Klebanov-Tseytlin
gauge theory compactified on $S^3$ \cite{buchelS3}. In
ref.\cite{Pons:2003ci} studies related to the Klebanov-Strassler
background have been carried out. However, it has been shown that using
Poincar\'e-like coordinates it is not possible to reproduce the
well-known field theory relation between energy and angular
momentum. The problems that one must face when studying near conformal
backgrounds are of two different types \cite{Tseytlin:2002ny}. One
concerns to the fact that near conformal backgrounds are written in
terms of Poincar\'e-like coordinates, while rotating string solutions
usually become simpler as written in global coordinates. The second
problem is related to the definition of the string theory analog of a
conformal dimension for near conformal backgrounds.

A complementary viewpoint in the context of type IIA string theory can
be developed under the observation that there is a map from certain
type IIA intersecting brane configurations onto the conifold. In fact,
time ago Bershadsky, Sadov and Vafa argued that the conifold
singularity is dual to a system of NS fivebranes intersecting over a
3+1 dimensional worldvolume \cite{Bershadsky:1995sp}.  Then, Dasgupta
and Muhki \cite{Dasgupta:1998su,Dasgupta:1999wx} have shown that a set
of parallel D3-brane probes near a conifold singularity can be mapped
onto a configuration of intersecting branes in type IIA string
theory. In this formulation they explicitly derived the field theory
on the probes. Moreover, it is possible to show that brane
constructions in type IIA string theory are related to large classes
of chiral and non-chiral ${\cal {N}}=1$ field theories with quartic
superpotentials \cite{uranga}. It would be of interest to have an
explicit realization for a system of type IIA intersecting branes
related to the deep IR limit of the Klebanov-Strassler
solutions.
In addition, a deformation of the conifold leads to a
smoothed out intersection of two sets of NS fivebranes
\cite{ohta}. Also, large $N$ dualities for a general class of ${\cal
{N}}=1$ theories on type IIB D5-branes wrapping 2-cycles of local
Calabi-Yau three-folds, or as seen as effective field theories on
D4-branes in type IIA brane configurations have been studied
\cite{Oh:2001bf}.

In this paper we investigate classical solutions in this complementary
framework leading to a new perspective about the dual description of
${\cal {N}}=1$ SYM theories in 4d.  We shall show how spinning and
rotating closed string solutions in certain type IIB string theory
backgrounds, which have been proposed to be dual to ${\cal {N}}=1$ SYM
theories, can be mapped onto classical string solutions embedded in
backgrounds induced by configurations of intersecting branes in type
IIA string theory. In the case of the conifold, for long string
solutions corresponding to large spin operators in the ultraviolet
limit of SYM theory, we found the expected logarithmic anomalous
dimension of minimal twist operators. In addition, infrared Regge
behavior for spinning short strings is recovered. In the type IIA
intersecting brane configurations studied here, there can also be
non-vanishing $B$-fields. We shall therefore study this effect that
provides new features of ${\cal {N}}=1$ SYM theory spectrum from the
brane construction perspective. Different solitonic string solutions
with $B$-fields were previously considered in
\cite{Rashkov:2002zt,alexandrova}.

In section 2 we review some results on classical solutions for
closed strings in type IIB string theory in conformal and
near-conformal backgrounds. In section 3 we study the
Dasgupta-Muhki map relating intersecting branes constructions to
the conifold. Then, in section 3.2 we obtain classical string
solutions in $AdS_5 \times T^{1,1}$ background, and we recover the
expected features for long and short spinning closed strings. We
map the above solutions onto classical solutions of rotating
closed strings embedded in type IIA intersecting brane
constructions.  In section 3.3 we study the specific case of
spinning closed strings in $AdS_5$ and rotating in $S^5$.  We
recover the expected features for long and short spinning closed
strings.  We also analyze the corresponding classical string
solutions in the intersecting brane constructions.  In section 4
we discuss about extensions of the map for the embedding of closed
string solutions in the UV limit of the Klebanov-Tseytlin
background. Then, using the Ohta-Yokono map we address similar
issues for the case of the deformed conifold. Inspired in this, we
also comment on classical solutions in the deep IR limit of the
Kebanov-Strassler background. Although we have not completely
succeeded in obtaining explicit relations of the form $E=E(S,J)$
for some of the string solutions presented in this paper, we think
this work can motivate to explore more exhaustively this kind of
relations in both, conifold-like and intersecting brane
backgrounds, and study the corresponding semi-classical
quantization, as well as string theory states/SYM theory operators
correspondence. The last section is devoted to a general
discussion, also describing interesting open questions related to
${\cal {N}}=1$ SYM theory operators, and M-theory descriptions of
the above mentioned systems.


\section{Rotating and spinning closed string solutions in type IIB string
theory in conformal and near-conformal backgrounds}

In this section we review some results of rotating and spinning
solutions for closed strings in type IIB string theory in
conformal backgrounds
\cite{Gubser:2002tv,Frolov:2002av,Tseytlin:2002ny}, that can be
extended to near-conformal ones \cite{Tseytlin:2002ny}.
Particularly, we review the case of boosted and spinning strings
in $AdS_5 \times S^5$ background. One of the reasons of interest
in this background is that for long strings the anomalous scaling
dimension of twist two operators behaves like $E=S+f(\lambda) \,
\ln S/\sqrt{\lambda} + \cdot \cdot \cdot$, where
$\sqrt{\lambda}=R^2/\alpha'$, being $R$ the $AdS_5$ radius. We
leave for section 3 a more extensive study of classical solutions
of $AdS_5 \times T^{1,1}$, and for section 4 a discussion on the
solutions related to non-conformal ${\cal {N}}=1$ SYM theories,
whose supergravity duals are the Klebanov-Tseytlin and
Klebanov-Strassler backgrounds.

Let us consider the bosonic part of the Green-Schwarz superstring
action in $AdS_5 \times S^5$
\be
I_B=- \frac{1}{2 \pi \, \alpha'} \, \int d\tau \, d\sigma \,\,
L_B(x),
\ee
where
\be
L_B(x)= \frac{1}{2} \, \sqrt{-g} \, g^{ab} \, G_{\mu\nu} \, \partial_a
x^\mu \, \partial_b x^\nu.
\ee
We use Minkowski signature in both, the target space and world-sheet. In
the conformal gauge we have $\sqrt{-g} \, g^{ab} = \eta^{ab} =
diag(-1,1)$.  In global coordinates the $AdS_5$ space-time metric can
be written as
\bea
ds^2_{AdS_5} &=& R^2 \, [-\cosh^2 \rho \,\, dt^2 + d\rho^2 + \sinh^2
\rho \,\, d\Omega_3] \, , \nonumber \\
d\Omega_3 &=& d\beta_1^2 + \cos^2\beta_1 \, (d\beta^2_2+\cos^2\beta_2
\, d\beta^2_3) \, , \label{globalAdSxS}
\eea
while the parametrization of $S^5$ is
\bea
ds^2_{S^5} &=& R^2 \, [d\psi_1^2 + \cos^2\psi_1 \,
(d\psi^2_2+\cos^2\psi_2 \, d\Omega'_3)] \, , \nonumber \\
d\Omega'_3 &=& d\psi_3^2 + \cos^2\psi_3 \,
(d\psi^2_4+\cos^2\psi_4 \, d\psi^2_5) \, , \label{globalS5}
%
%
\eea
Then, for a closed spinning string in $\phi \equiv \beta_3$ of $AdS_5$
and boosted along the direction $\varphi \equiv \psi_5$ of $S^5$, one
can write the ansatz
\bea
&& t= \kappa \, \tau, \,\,\,\,\,\,\,\, \phi = \omega \, \tau,
\,\,\,\,\,\,\,\, \varphi = \nu \, \tau, \nonumber \\
&& \rho(\sigma) = \rho(\sigma+2\pi), \,\,\,\,\,\,\,\,\, \beta_i = 0
\,\,\,\,\,\, (i=1, \, 2), \,\,\,\,\,\,\,\,\, \psi_j=0 \,\,\,\,\,\,
(j=1, \cdot \cdot \cdot 4),
\eea
where $\kappa$, $\omega$ and $\nu$ are constants. $\rho$ is subject to
the following second order equation
\be
\rho'' = (\kappa^2 - \omega^2) \, \cosh\rho \, \sinh\rho,
\label{secondderivative}
\ee
which is implied by Eq.(\ref{rhoprima}) below,
where prime stands for derivative with respect to $\sigma$.
The conformal constraints are
\be
G_{\mu\nu} \, \frac{\partial X^\mu}{\partial \tau} \, \frac{\partial
X^\nu}{\partial \sigma} = 0 , \,\,\,\,\,\,\,\,\,\,\,\,\,\
G_{\mu\nu} \, \left( \frac{\partial X^\mu}{\partial \tau} \, \frac{\partial
X^\nu}{\partial \tau} +  \frac{\partial X^\mu}{\partial \sigma} \, \frac{\partial
X^\nu}{\partial \sigma} \right) = 0 .
\ee
The first equation is automatically satisfied for the above string
configuration.  Using the second conformal constraint and the
previous ansatz for the string solution, it is derived the
following relation
\be
(\rho')^2 = \kappa^2 \, \cosh^2\rho - \omega^2 \, \sinh^2\rho - \nu^2,
\label{rhoprima}
\ee
>From Eq.(\ref{rhoprima}) one straightforwardly obtains
\be
2 \, \pi = \int_0^{2 \pi} \, d\sigma =
4 \, \int_0^{\rho_0} \, d\rho \, \frac{1}{\sqrt{(\kappa^2-\nu^2) \,
\cosh^2\rho - (\omega^2 - \nu^2) \, \sinh^2\rho}}.
\ee
There are three conserved quantities, {\it i.e.} energy, spin and
R-symmetry charge $J$,
\bea
E&=&\sqrt{\lambda} \, \kappa \, \int_0^{2\pi} \frac{d\sigma}{2 \pi} \,
\cosh^2\rho \equiv \sqrt{\lambda} \, {\cal {E}}, \\
S&=&\sqrt{\lambda} \, \omega \, \int_0^{2\pi} \frac{d\sigma}{2 \pi} \,
\sinh^2\rho \equiv \sqrt{\lambda} \, {\cal {S}}, \\
J&=&\sqrt{\lambda} \, \nu \, \int_0^{2\pi} \frac{d\sigma}{2 \pi}
\equiv \sqrt{\lambda} \, \nu.
\eea
Thus, one obtains the relation
\be
E=\frac{\kappa}{\nu} \, J + \frac{\kappa}{\omega} \, S. \label{eeq12}
\ee
At this point, one can explicitly calculate the above integrals in
terms of hyper-geometric functions. It is useful to write down the
energy-spin relations for both, short and long strings. In doing this,
it is convenient to introduce a parameter $\eta > 0$, defined as
\be
\coth^2 \rho_0 = \frac{\omega^2 - \nu^2}{\kappa^2 - \nu^2} = 1 + \eta
. \label{etaeta}
\ee
In fact, for short strings Eq.(\ref{eeq12}) becomes
\be
{\cal {E}} \approx \sqrt{\nu^2 + \frac{2 \, {\cal
{S}}}{\sqrt{1+\nu^2}}} + \frac{\sqrt{\nu^2 + \frac{2 \, {\cal
{S}}}{\sqrt{1+\nu^2}}}} {\sqrt{1 + \nu^2 + \frac{2 \, {\cal
{S}}}{\sqrt{1+\nu^2}}}} \,\, {\cal {S}} .
\label{ES}
\ee
The above expression is valid when $\frac{1}{\eta} \approx \frac{2
\, {\cal {S}}}{\sqrt{1+\nu^2}} << 1$. Now, if we demand $\nu<<1$,
then ${\cal {S}} << 1$, and therefore Eq.(\ref{ES}) reduces to
\be
E^2 \approx J^2 + 2 \, \sqrt{\lambda} \, S.
\ee
This expression is the limit for short strings spinning and rotating
in $AdS_5 \times S^5$. They probe a small curvature region of
$AdS_5$. Moreover, if the boost energy is much smaller than the
rotational energy, {\it i.e.} $\nu^2<<{\cal {S}}$, then
\be
{\cal {E}} \approx \sqrt{2 \, {\cal {S}}} +
\frac{\nu^2}{2 \, \sqrt{ 2 \, {\cal {S}}}} ,
\ee
which indeed is the flat-space Regge trajectory. Besides, when the
boost energy is greater than the spin ($2 {\cal {S}}<<\nu$) one
obtains
\be
E \approx J + S + \frac{\lambda \, S}{2 \, J^2}.
\ee
On the other hand, for long strings, {\it i.e.} $\eta \rightarrow
0^{+}$, for $\nu << - \log \eta$, it is obtained the relation
\be
E \approx S + \frac{\sqrt{\lambda}}{\pi} \, \log (S/\sqrt{\lambda}) +
\frac{\pi J^2}{2 \sqrt{\lambda} \log (S/\sqrt{\lambda})} .
\ee
For $\nu = 0$, it becomes the relation that, through the holographic
identification $E \equiv \Delta$ (where $\Delta$ is the scaling
dimension of the corresponding SYM theory operator), leads to the
logarithmic anomalous dimension of minimal twist operators.

When $\log ({\cal {S}}/\nu) << \nu << {\cal {S}}$, we obtain
\be
E \approx J + S + \frac{\lambda}{2 \, \pi^2 \, J} \, \log^2 (S/J) .
\ee
Now, it is also useful to review some issues given in
refs. \cite{Gubser:2002tv,Tseytlin:2002ny}, relating results
expressed in global and Poincar\'e coordinates.  In Poincar\'e
coordinates the metric of $AdS_5 \times S^5$ space-time is
\be
ds^2_{AdS_5 \times S^5} = \frac{R^2}{z^2} \, (dx_m dx_m + dz_p
dz_p) \, ,
\ee
where $m=0, \cdot \cdot \cdot , 3$ and $p=1, \cdot \cdot \cdot , 6$,
while $z_p z_p =z^2$.
Let us consider $z=z(\tau, \sigma)$ as the radial coordinate in the
Poincar\'e parametrization.  The transformations between Poincar\'e
and global coordinates of anti de Sitter space-time are the following
\bea
X_0 &=& \frac{x_0}{z} = \cosh\rho \, \sin t \, , \nonumber \\
X_i &=& \frac{x_i}{z} = n_i \, \sinh\rho \, , \nonumber \\
X_4 &=& \frac{1}{2 z} (-1+z^2-x_0^2+x_i^2) = n_4 \sinh\rho \, ,
\,\,\,\,\,\, n_i^2+n_4^2 =1 \, , \nonumber
\\
X_5 &=& \frac{1}{2 z} (1+z^2-x_0^2+x_i^2) = \cosh\rho \, \cos t \,
\nonumber
\\
\tan t &=& \frac{2 x_0}{1+z^2-x_0^2+x_i^2} \, , \nonumber \\
z &=& \frac{1}{\cosh\rho \, \cos t - n_4 \, \sinh\rho} \, .
\eea
The variables $X_0, X_i, X_4$ and $X_5$ are the coordinates of
$\mathbb{R}^{2,4}$, being the metric of the $AdS_5$ induced from the
flat $\mathbb{R}^{2,4}$ metric by the embedding
$X_0^2+X_5^2-X_i^2-X_4^2=1$. The unit vector $n_k$ parametrizes the
$S^3$ in $AdS_5$.

The relation between the energy of $AdS_5$ expressed in Poincar\'e
and global coordinates is given by
\bea
E &=& \sqrt{\lambda} \, {\cal {E}} =  \frac{\sqrt{\lambda}}{2 \pi}
\, \int d\sigma \, {\cal {E}}_d =  \frac{\sqrt{\lambda}}{2 \pi} \,
\int d\sigma \, \dot t \, \cosh^2\rho \nonumber \\
&=& \frac{\sqrt{\lambda}}{4 \, \pi} \, \int d\sigma \,
[(1+z^2+x^2) {\cal {P}}_0- 2 \, x_0 \, {\cal {D}}] \, .
\label{energy100}
\eea
%
Energy density for translations in $x_0$ is
\be
{\cal {P}}_0 = \frac{1}{z^2} \, \dot {x}_0 \, ,
\ee
and dilatation charge density is given by
\be
{\cal {D}} = \frac{1}{2 z^2} \, \frac{\partial}{\partial \tau}
(z^2+x^2) \, ,
\ee
with $x^2=-x^2_0+x_i^2$. In addition, in the Poincar\'e patch
\be
E = \frac{P_0+K_0}{2}= \frac{\sqrt{\lambda}}{4 \pi} \int d\sigma
({\cal {P}}_0+{\cal {K}}_0),
\ee
where ${\cal {K}}_0=(z^2+x^2) {\cal {P}}_0 - 2 x_0 {\cal {D}}$.
The energy density in Eq.(\ref{energy100}) results
\be
{\cal {E}}_d = \frac{(1+z^2+x^2)^2}{2 z^2} \,
\frac{\partial}{\partial \tau} \left( \frac{x_0}{1+z^2+x^2}
\right) \, . \label{energy-AdS}
\ee
In particular, for the Klebanov-Tseytlin background a very
interesting solution has been obtained \cite{Tseytlin:2002ny},
derived from a deformation of a point-like string boosted along
the circle parametrized by $\varphi$ on $S^5$. In Poincar\'e
coordinates we parametrize the undeformed solution as
\be
x_0=\tan t \, , \,\,\,\,\,\, z=\frac{1}{\cos t} \, , \,\,\,\,\,\,
\varphi=t=\nu \, \tau \, , \label{boost-poincare}
\ee
while, in global coordinates it reads
\be
t = \nu \, \tau \, , \,\,\,\,\,\, \varphi=\nu \, \tau \,
. \label{boost-global}
\ee
Similarly, a deformation of a closed string spinning in $AdS_5$ has
been considered \cite{Tseytlin:2002ny}.  In Poincar\'e coordinates the
undeformed solution is
\be
x_0=\tan t \, , \,\,\,\,\,\, z=\frac{1}{\cos t \, \cosh\rho} \, ,
\,\,\,\,\,\, x_1=r\, \cos\phi \, , \,\,\,\,\,\, x_2=r \, \sin\phi
\, , \,\,\,\,\,\, r = \frac{\tanh\rho}{\cos t} \, ,
\label{rotating-poincare}
\ee
and in global coordinates it becomes
\be
 t = \kappa \, \tau \, , \,\,\,\,\,\, \phi=\omega \, \tau \, ,
 \,\,\,\,\,\, \rho=\rho(\sigma) \, , \,\,\,\,\,\, \rho'^2=\kappa^2 \,
 \cosh^2\rho - \omega^2 \, \sinh^2\rho. \label{rotating-global}
\ee
For this kind of solutions the energy $E$ in global coordinates
coincides with the energy in the Poincar\'e ones, $P_0$. This is so
because $z^2-x_0^2+x_i^2=1$.

In addition, for certain solutions in which we shall be
interested, on the side of the type IIA brane constructions we are
going to deal with configurations including non-vanishing
$B$-fields. In some cases, this will introduce certain corrections
to the spectra of spinning and rotating strings. For example, they
will modify some energy-spin and energy-R symmetry charge relation
where the rotating strings are stretched along $\theta_i$
direction in $T^{1,1}$. So, in this situation $\theta_i$ will be a
function of $\sigma$, and the bosonic Lagrangian will be
\be
I = I_B - \frac{1}{4 \, \pi \, \alpha'} \, \int \, d\tau \, d\sigma \,
\epsilon^{ab} \, B_{\mu\nu} \, \partial_a x^\mu \, \partial_b x^\nu,
\ee
where $\epsilon^{ab}$ is the Levi-Civita tensor density, and
$B_{\mu\nu}$ is the Neveu-Schwarz anti-symmetric $B$-field.
However, we shall see that for the configurations studied here the
$B$-field term in the action vanishes. Therefore, the only
$B$-field contributions that will be present in the calculation of
energy, spin as well as R-symmetry charge, come from the
Nambu-Goto Lagrangian, {\it i.e.} from the type IIB metric
obtained after T and S-duality transformations are applied on the
NS5-NS5'-D4 configuration, as we shall see explicitly in the next
section. In $AdS_5 \times S^5$ background non-vanishing $B$-fields
configurations have been studied in references
\cite{Rashkov:2002zt,alexandrova}.


\section{Brane cons\-truc\-tions in type IIA and type IIB string theo\-ries and
cla\-ssi\-cal so\-lu\-tions for clo\-sed strings}

In this section we firstly review the Dasgupta-Mukhi construction
which maps a configuration of intersecting branes in type IIA
string theory onto a type IIB configuration of $N$ parallel
D3-branes on the conifold singularity. Then, inspired on this
construction we map classical closed string solutions in $AdS_5
\times T^{1,1}$ background onto classical solutions of strings
embedded in configurations of intersecting branes in type IIA
string theory.

\subsection{Brane constructions and the conifold}

The construction proposed by Dasgupta and Mukhi (DM) uses a
version of an earlier brane construction developed by Hanany and
Witten \cite{Hanany:1996ie}. DM construction enables one to
read off the spectrum and other properties of the conformal field
theory on the D3-branes world-volume at the conifold. DM argue
that the conifold singularity is represented by a configuration of
two type IIA NS 5-branes rotated with respect to each other, and
located on a circle $S^1$, with D4-branes stretched between the NS
5-branes from both sides along $S^1$.

Let us start by considering a type IIB configuration of a D3-brane
stretched between two D5-branes perpendicular to each other, and
separated along $x^6$ (this is the compact direction that parametrizes
$S^1$) as indicated below
\bea
D5\,&& 0 \,\,\, 1 \,\,\, 2 \,\,\,\, 3 \,\,\,\, 4 \,\,\, 5 \, -  -
- - \nonumber
\\
D5'&& 0 \,\,\, 1 \,\,\, 2 \,\,\,\, 3 \,\, -  -  -  - \,\, 8 \,\,\,
9 \nonumber
\\
D3\,&& 0 \,\,\, 1 \,\,\, 2 \,\,\, -  -  -  \, 6 \, -  - \, -
\nonumber
\eea
In order to obtain an explicit expression for the metric of such a
configuration we use the rules give in references
\cite{Bergshoeff:1995as,Tseytlin:1996bh,
Tseytlin:1997cs,Gauntlett:1997cv,Andreas:1998hh,Ohta03}, {\it
i.e.} given $M$ intersecting Dp-branes with harmonic functions
$H_p$, one must choose the maximal set of common directions $n_1$.
The metric for that piece has a common factor $(H_{p_1} \, H_{p_2}
\, \cdot \cdot \cdot \, H_{p_{m_1}})^{-1}$, where $m_1$ is the
number of Dp-branes with $n_1$ common directions. In the present
case $n_1$ is $3$, corresponding to $x^0, \, x^1$ and $x^2$, and
leading to the factor $(H_3 \, H_5 \, H'_5)^{-1}$ for the piece
$ds^2_{012}$. Then, we must proceed in a similar way with the
second set of common directions. In this case, this is given by
the direction $x^3$, including the factor $(H_5 \, H'_5)^{-1}$. In
this way, the metric becomes
\bea
ds^2 & = &(H_3 \, H_5 \, H'_5)^{1/2} \, [(H_3 \, H_5 \, H'_5)^{-1}
\, ds^2_{012} + (H_5 \, H'_5)^{-1} \, ds^2_3 + (H_5)^{-1} \,
ds^2_{45} \nonumber  \\
& & + (H_3)^{-1} \, ds^2_6 + ds^2_7 + (H_5')^{-1} \, ds^2_{89} ]
\,\,\, ,
\eea
where we see that there is an overall factor. The harmonic functions
$H_3, \, H_5$ and $H'_5$ only depend on a single overall transverse
coordinate $x^7$. Thus, all $H_i$ behave like $1+|x^7|$. Thus,
we have a configuration of partially intersecting branes
\cite{Tseytlin:1996bh,Tseytlin:1997cs,youm,ohta,hashimoto}.

Next, we study the metric above under S-duality transformations
leading to two perpendicular NS 5-branes as follows
\bea
NS 5\,&& 0 \,\,\, 1 \,\,\, 2 \,\,\,\, 3 \,\,\,\, 4 \,\,\, 5 \, -
- - - \nonumber
\\
NS 5'&& 0 \,\,\, 1 \,\,\, 2 \,\,\,\, 3 \,\, -  -  -  - \,\, 8
\,\,\, 9 \nonumber
\\
D3\,\,\,\,\,\,&& 0 \,\,\, 1 \,\,\, 2 \,\,\, -  -  -  \, 6 \, -  -
\, - \nonumber
\eea
being the S-dual metric obtained by multiplying the previous one by an
overall factor $(H_5 \, H'_5)^{1/2}$
\bea
ds^2 & = &(H_3)^{-1/2} \, ds^2_{012} + (H_3)^{1/2} \, ds^2_3 +
(H_3)^{1/2} \, H'_5 \, ds^2_{45} + (H_3)^{-1/2} \, H_5 \, H'_5 \,
ds^2_6  \nonumber  \\
& & +(H_3)^{1/2} \, H_5 \, H'_5 \, ds^2_7 + (H_3)^{1/2} \, H_5 \,
ds^2_{89} \,\,\, .
\eea
Now, we obtain the T-dual version of the metric above in the
direction $x^3$, so that we get a type IIA metric corresponding to
D4-branes stretched between the perpendicular NS 5-branes
\bea
ds^2 & = &(H_3)^{-1/2} \, ds^2_{0123} + (H_3)^{1/2} \, H'_5 \,
ds^2_{45} + (H_3)^{-1/2} \, H_5 \, H'_5 \, ds^2_6 \nonumber  \\
& &  +(H_3)^{1/2} \, H_5 \, H'_5 \, ds^2_7 + (H_3)^{1/2} \, H_5 \,
ds^2_{89} \,\,\, ,
\eea
while the intersecting brane configuration is
\bea
NS 5\,&& 0 \,\,\, 1 \,\,\, 2 \,\,\,\, 3 \,\,\,\, 4 \,\,\, 5 \, - -
- - \nonumber
\\
NS 5'&& 0 \,\,\, 1 \,\,\, 2 \,\,\,\, 3 \,\, -  -  -  - \,\, 8
\,\,\, 9 \nonumber
\\
D4\,\,\,\,\,\,&& 0 \,\,\, 1 \,\,\, 2 \,\,\,\, 3 \,\,  -  -  \,\, 6
\, - - \, - \nonumber
\eea
At this point, we use the duality map providing the relation between
the metric in type IIA string theory, $g$, and the one in type IIB
string theory, $G$, (see
refs.\cite{Bergshoeff:1995as,Dasgupta:1998su}). Notice that on the
type IIA string theory configuration non-trivial $B$-fields can be
expected. Thus, there are the following relations
\be
G_{mn} = g_{mn} - (g_{6m} \,  g_{6n} - B_{6m} \, B_{6n}) \,
g^{-1}_{66} \, , \,\,\,\,\,\,\,\, G_{66}=g^{-1}_{66} \, ,
\,\,\,\,\,\,\,\,\, G_{6m} = B_{6m} \, g^{-1}_{66} \, ,
\ee
where $m$ and $n$ take values from 0 to 9, except 6. In this case
we assume non-zero values of $B$-fields only for the
components $B_{64}$ and $B_{68}$, so that
\bea
G_{\mu \nu} & = & g_{\mu \nu}, \,\,\,\,\,\, \mu, \nu = 0, 1, 2, 3,
\nonumber \\ G_{66} & = & g^{-1}_{66}, \nonumber \\ G_{6i} & = &
B_{6i} \, g^{-1}_{66}, \,\,\,\,\,\, i=4, 8, \nonumber \\ G_{ii} & = &
g_{ii} + B^2_{6i} \, g^{-1}_{66}, \nonumber \\ G_{77} & = & g_{77},
\nonumber \\ G_{48} & = & B_{64} \, B_{68} \, g^{-1}_{66}.
\eea
Therefore, the new type IIB metric is
\bea
ds^2 & = &(H_3)^{-1/2} \, ds^2_{0123} + (H_3)^{1/2} \, [ H'_5 \,
ds^2_{45} +  H_5 \, H'_5 \, ds^2_7 + H_5 \, ds^2_{89} \nonumber
\\
& &  + (H_5 \, H'_5)^{-1} \, (B_{64} \, ds_4 + B_{68} \, ds_8 +
ds_6)^2] \,\,\, . \label{metricIIBnew}
\eea
Following DM we shall argue that this brane configuration is locally a
D3-brane at the conifold singularity. This argument is based on the
following facts. Firstly, one must consider $H_3$ to be the harmonic
function of a D3-brane localized in the 6 transverse coordinates
$x_i$, $i=4$ to $9$. On the other hand, $H_5$ and $H'_5$ will
continue to be as before $1+|x^7|$. It means that both 5-brane
harmonic functions are non-singular and they approach a constant when
$x^7 \rightarrow 0$. This allows one to eliminate the 5-brane harmonic
functions through a re-scaling of the coordinates $(x^4, x^5)$ and
$(x^8, x^9)$, respectively. Finally, it is worth to remark that
although the directions $(x^4, x^5)$ and $(x^8, x^9)$ constitute two
planes, we need to combine them into the direct product of two
2-spheres with definite radius. This implies an enhancement of
global symmetry $U(1) \times U(1)$ to $SU(2) \times SU(2)$.

In order to complete the map we write down certain necessary
redefinitions. Let us define $\omega_4 = B_{64}$, and similarly
$\omega_8 = B_{68}$, and determine them by solving ${\vec
{\nabla}} \times {\vec \omega} = const.$ By defining $\vec
\omega_i = (\omega_i, 0, 0 )$ for $i=4$ and 8, and using the $curl$ in
polar coordinates we get the differential equation
\be
\frac{1}{\sin\theta_1} \, \frac{\partial(\sin\theta_1 \, \omega_4
)}{\partial\theta_1} = const = {\tilde A_1} \,\,\, .
\ee
We can solve it obtaining the solution $\sin\theta_1 \, \omega_4 =
- \cos\theta_1 \, {\tilde A_1} + {\tilde A_2}$. Particularly, we
can set the constants to be ${\tilde A_1}=-D$, while ${\tilde
A_2}=0$. Therefore, we get $\omega_4 = D \, \cot\theta_1$ and
$\omega_8 = D \, \cot\theta_2$, respectively. In addition, as we
have already remarked the fact that the metric
(\ref{metricIIBnew}) resembles the structure of D3-branes at the
conifold singularity is an encouraging signal in order to attempt
to transform that metric in the conifold one. Through the
following redefinitions (here we will use the notation $dx^i
\equiv ds_i$)
\be
ds_4 \rightarrow \sqrt{C} \, \sin\theta_1 \, d\phi_1 \, ,
\,\,\,\,\,\,\, ds_5 \rightarrow \sqrt{C} \, d\theta_1 \, ,
\label{x4}
\ee
and
\be
ds_8 \rightarrow \sqrt{C} \, \sin\theta_2 \, d\phi_2 \, ,
\,\,\,\,\,\,\, ds_9 \rightarrow \sqrt{C} \, d\theta_2 \, ,
\label{x8}
\ee
we get the transformation
\be
ds^2_{45} + ds^2_{89} \rightarrow C \, \sum_{i=1}^2 (d\theta_i^2 +
\sin\theta_i^2 \, d\phi_i^2) \,\,\, .
\ee
In addition, we identify the direction $x^6$ on the metric
(\ref{metricIIBnew}) with the Hopf fiber $\psi$ multiplied by the
integration constants $\sqrt{C} \, D$. We have assumed that $x^6$
is compactified on $S^1$ while, on the other hand, the Hopf fiber
takes values in the range $[0, 4 \pi)$. Including all the
mentioned replacements we get
\bea
(B_{64} \, ds_4 + B_{68} \, ds_8 + ds_6)^2 \rightarrow C D^2
(d\psi + \cos\theta_1 \, d\phi_1 + \cos\theta_2 \, d\phi_2)^2
\,\,\, ,
\eea
setting $C D^2=1/9$ and $C=1/6$, using the redefinition $x_7=\log
r$ and suitably rescaling coordinates, we get the metric
\bea
ds^2 & = &(H_3)^{-1/2} \, ds^2_{0123} + (H_3)^{1/2} \, [ dr^2 +
r^2 \, \frac{1}{6} \, \sum_{i=1}^2 (d\theta_i^2 + \sin\theta_i^2
\, d\phi_i^2) \nonumber \\
& & + r^2 \, \frac{1}{9} \, (d\psi + \cos\theta_1 \, d\phi_1 +
\cos\theta_2 \, d\phi_2)^2] \,\,\, , \label{conifoldmetric}
\eea
which turns out to be the metric of D3-branes at the conifold
singularity.

At this point one interesting question arises, it is whether we
can follow the DM construction in a reverse way, {\it i.e.}
starting from the conifold metric (\ref{conifoldmetric}). An
interesting point consists in breaking the $SU(2) \times SU(2)$
global symmetry down to $U(1) \times U(1)$ global symmetry. In
doing this we must guarantee that the base of the conifold $S^2
\times S^3$ will be mapped onto two 2-planes plus a circle which
will be related to the compact $x^6$ coordinate in the brane
construction. The first point to notice is that the piece defining
the infinitesimal length on any of $S^2$'s
\be
d\theta^2_i + \sin^2\theta_i \, d\phi^2_i , \label{polarmetric}
\ee
for small values of $\theta_i$ looks like
\be
d\theta^2_i + \theta^2_i \, d\phi^2_i \,\,\, .
\ee
This is nothing but the manifestation of the elementary fact that
we can, at least locally, break the isometry $SU(2)$ down to
$U(1)$. So, we now can take $\theta_i$ to be a radial variable
${\tilde {\rho}}$ (in the sense of polar coordinates), and by
extending the metric (\ref{polarmetric}) for all positive values
of ${\tilde {\rho}}$ we get the inversion of the relations
(\ref{x4}) and (\ref{x8}). In the next section we start discussing
the rest of transformations in order to get the brane
constructions for the specific cases in which we are interested.

\subsection{Classical solutions for closed strings in type IIA and
type IIB backgrounds}

In reference \cite{Gubser:2002tv} it has been obtained classical
closed string solutions rotating around the equatorial circle of
$S^5$, as well as, in the three-sphere parametrizing the anti de
Sitter space, and it has been understood how these solutions are
related to large R-symmetry charge and large spin operators in the
dual ${\cal {N}}=4$ $SU(N)$ SYM theory in 4d, respectively. For our
present program we are firstly interested in four kinds of closed
string solutions rotating and spinning in the conifold background. The
first one corresponds to a closed point-like string orbiting along
$\psi$.  Since after undoing the DM construction the Hopf fiber is
related to the compact $x^6$ in the intersecting brane construction,
this solution will be mapped onto a point-like closed string orbiting along
$S^1$.

The second solution corresponds to a point-like closed string
boosted along $\theta_i$ in a 2-cycle of $T^{1,1}$.  This solution
maps onto a point-like string boosted along the direction $x^4$
(or $x^8$) in the type IIA brane setup. The third kind of solution
that we study here is a closed string centered in any of the two
2-spheres in $T^{1,1}$, and rotating around its center. When it is
transformed onto the corresponding embedding in the intersecting
brane construction, this configuration renders two types of
equivalent classical closed string solutions moving on the planes
$(x^4, x^5)$ and $(x^8, x^9)$, respectively. We also study the
case corresponding to a closed string rotating on the $S^3$ in
$AdS_5$, that can be understood as a closed string rotating in the
directions $x^1, x^2, x^3$ on the type IIA intersecting brane
construction.

We start from the near horizon limit of the metric
(\ref{conifoldmetric}). Rewriting it global coordinates
\bea
ds^2 & = & L^2 \, [-\cosh^2\rho \, dt^2 + d\rho^2 + \sinh^2\rho \,
d\Omega_3 +  \frac{1}{6} \, \sum_{i=1}^2 (d\theta_i^2 +
\sin\theta_i^2 \, d\phi_i^2) \nonumber
\\
& & + \frac{1}{9} \, (d\psi + \cos\theta_1 \, d\phi_1 +
\cos\theta_2 \, d\phi_2)^2] \,\,\, , \label{conifoldmetricglobal}
\eea
where $L^4= \left( \frac{27}{16} \right) \, 4 \, \pi \, g_s \, N \,
\alpha'^2$.

~

~

{\it A point-like closed string boosted along the fiber $\psi$ in
$T^{1,1}$}

~

Consider the following parameterization for a string solution
embedded in the conifold metric given by
\bea
&&t= \kappa \, \tau, \,\,\,\,\,\,\,\, \psi=\nu \, \tau, \nonumber \\
&&\rho=0, \,\,\,\,\,\,\,\, \theta_i=\phi_i=0, \,\,\,\, i=1, 2,
\,\,\,\,\,\,\,\, \beta_{1,2,3}=0,
\eea
where $\beta_i$ belongs to $\Omega_3$. Therefore, using the Nambu-Goto
Lagrangian the relation between energy and $U(1)$ R-symmetry charge is
given by
\be
E  =  \frac{9 \, \kappa}{\nu} \, J .
\label{boosted}
\ee
Henceforth we take $\sqrt{\lambda}=L^2/\alpha'$, {\it i.e.} using the
conifold radius instead of the $S^5$ one. Since we are interested in
studying how this $E-J$ relation changes when we map the conifold onto
the NS5-NS5'-D4 brane system, it is useful to perform the following
identifications
\bea
&&\frac{\psi}{3} \equiv \frac{\nu \, \tau}{3} \equiv x_6, \nonumber \\
&&\sqrt{C} \, \sin\theta_j \, d\phi_j \rightarrow dx_i, \,\,\,\,\,\,\,\,\,
j=1, 2, \,\,\,\,\, i=4, 8, \nonumber \\
&&\sqrt{C} \, d\theta_j \rightarrow  dx_k , \,\,\,\,\,\,\,\,\,
j=1, 2, \,\,\,\,\, k=5, 9, \label{map1}
\eea
so that in this way we can write down the metric as seen by the closed
point-like string boosted along $x^6$ direction, {\it i.e.} using
Eq.(\ref{metricIIBnew})
\be
ds^2 = - H_3^{-1/2} \, dt^2 + H_3^{1/2} \, (H_5 \, H'_5)^{-1} \, dx_6^2.
\ee
This is a type IIB metric related to the conifold one through the map
(\ref{map1}). Here we have parametrized the corresponding string
solution as $t = \kappa \, \tau$, $x^6 = \nu \tau/3$, and $x^i=0$ for
$i=1, \cdot  \cdot  \cdot , 5$, 8, and 9. We take $x^7$ to be a constant.

Now, we use the relation between the metric, $G$, in type IIB string
theory and the one in type IIA string theory $g$, {\it i.e.},
$G_{00}=g_{00}, \,\,\, G_{66}=g^{-1}_{66}=H_3^{1/2} \, (H_5 \,
H'_5)^{-1}$. Therefore, the type IIA metric reads
\be
ds^2 = - H_3^{-1/2} \, dt^2 + H_3^{-1/2} \, H_5 \, H'_5 \, dx_6^2 ,
\ee
being the solution a point-like closed string boosted along the
compact direction $x^6$ as above. One can easily calculate the
relation
\be
E  =  \frac{1}{H_5 \, H'_5} \, \frac{9 \, \kappa}{\nu} \, J .
\label{boosted-IIA}
\ee
Furthermore, one can apply T-duality in direction $x^3$, and then
using S-duality transformations one gets a boosted string along $x^6$,
but in the system of $N$ D3-branes stretched along two perpendicular
D5-branes.  The energy and $J$ become related again through
Eq.(\ref{boosted-IIA}), that in the limit for $|x^7| \rightarrow 0$
reduces to relation (\ref{boosted}).

~

~

{\it A point-like closed string boosted along $\theta_i$ in
a 2-cycle of $T^{1,1}$}

~

One can also consider multi-spin solutions as in
ref.\cite{Frolov:2003qc}.  In this paper we study the case of a
single-folded closed string only spinning in one angular
direction.  For a string embedded in the conifold metric we use
the ansatz
\bea
&& t= \kappa \, \tau, \,\,\,\,\,\,\,\, \phi_1=\omega_1 \, \tau,
\,\,\,\,\,\,\,\, \psi=0, \nonumber \\ && \sin\theta_i = a_i,
\,\,\,\,\,\,\,\, \cos\theta_i = b_i, \,\,\,\, i=1, \, 2, \nonumber
\\ && \rho=0, \,\,\,\,\,\,\,\, \phi_2 = constant, \,\,\,\,\,\,\,\,
\beta_{1,2,3}=0 .
\eea
Therefore, the metric as seen by the string is
\be
ds^2 = L^2 [-\kappa^2 \, d\tau^2 + \frac{1}{6} \, a_1^2 \, \omega_1^2
 \, d\tau^2 + \frac{1}{9} \, b_1^2 \, \omega_1^2 \, d\tau^2] .
\ee
The Nambu-Goto Lagrangian and action read
\be
{\cal {L}}_{NG}=L^2 \, \left( \kappa^2-\omega^2_1 \, \left(
\frac{a^2_1}{6}+\frac{b^2_1}{9} \right) \right)^{1/2}, \,\,\,\,
\,\,\,\, I_{NG}=-\frac{\sqrt{\lambda}}{2 \pi} \, \int \, d\sigma \,
d\tau \, {\cal {L}}_{NG} .
\ee
Energy and $U(1)$ R-symmetry charge are given by
\bea
E & = & - \frac{\partial {\cal {L}}_{NG}}{\partial \kappa} = \kappa \,
\frac{\sqrt{\lambda}}{2 \pi} \, \int_0^{2 \pi} d\sigma
\frac{1}{\sqrt{\kappa^2 -\omega^2_1 \, \left(
\frac{a^2_1}{6}+\frac{b^2_1}{9} \right) }} , \nonumber \\
J_1 & = & \frac{\partial {\cal {L}}_{NG}}{\partial \omega_1} =\left(
\frac{a^2_1}{6}+\frac{b^2_1}{9} \right)
\omega_1 \, \frac{\sqrt{\lambda}}{2 \pi} \, \int_0^{2 \pi} d\sigma
\frac{1}{\sqrt{\kappa^2 -\omega^2_1 \, \left(
\frac{a^2_1}{6}+\frac{b^2_1}{9} \right) }} ,
\eea
and they are related by
\be
E = \frac{\kappa \, \, J_1}{\omega_1 \, \left(
\frac{a^2_1}{6}+\frac{b^2_1}{9} \right)} . \label{EK11}
\ee
Now, let us identify $\sqrt{C} \, \sin\theta_1 \, d\phi_1
\equiv 1/\sqrt{6} \, a_1 \, \omega_1 \, d\tau \rightarrow dx^4 $,
where $C=1/6$, while $x^6 = 0$ and $dx_5=dx_8=dx_9=0$. Thus, the type
IIB metric (\ref{metricIIBnew}) as seen by the string is
\be
ds^2 = -H_3^{-1/2} \, dt^2 + H_3^{1/2} \, [H'_5 + (H_5 H'_5)^{-1} \,
B_{64}^2] \, dx^2_4 , \label{metric58}
\ee
where we see that a non-trivial $B$-field is turned on. This is
$B_{64} = \sqrt{2/3} \, \cot\theta_1 = \sqrt{2/3} \, b_1/a_1$.
Since we have taken $x^6$ as a constant in the metric
(\ref{metric58}), the second term in the action (32) vanishes.
However, they do contribute to the energy and momentum through the
term $ \frac{2}{3} \, \frac{b^1_1}{a^2_1} \, (H_5 H'_5)^{-1}$. The
conjugate momentum related to $x^4$, $J_1$, corresponds to a
point-like string moving along $x^4$ direction. We can write again
the Nambu-Goto Lagrangian and get
\bea
E &=& - \frac{\partial {\cal {L}}_{NG}}{\partial \kappa} = \kappa \,
\frac{1}{2 \pi \alpha'} \, \int_0^{2 \pi} d\sigma
\frac{H_3^{-1/2}}{\sqrt{H_3^{-1/2} \, \kappa^2 - \frac{1}{6} \, a^2_1
\, \omega^2_1 \, H_3^{1/2} \, \left(H'_5 + \frac{2}{3} \,
\frac{b^1_1}{a^2_1} \, (H_5 H'_5)^{-1} \, \right) }} , \nonumber \\
J_1 &=& \frac{\partial {\cal {L}}_{NG}}{\partial \omega_1} =
\frac{\omega_1 \, a^2_1}{6} \, \frac{1}{2 \pi \alpha'} \, \int_0^{2
\pi} d\sigma \frac{H_3^{1/2} \, (H'_5 + \frac{2}{3} \,
\frac{b^1_1}{a^2_1} \, (H_5 H'_5)^{-1} \,
)}{\sqrt{H_3^{-1/2} \, \kappa^2 - \frac{1}{6} \,
a^2_1 \, \omega^2_1 \, H_3^{1/2} \, \left(H'_5 + \frac{2}{3} \,
\frac{b^1_1}{a^2_1} \, (H_5 H'_5)^{-1} \right) }} . \nonumber \\
&&
\eea
Therefore, we get the relation
\be
E = \frac{6 \, \kappa \, J_1}{a^2_1 \, \omega_1 \, H_3 \, \left(H'_5 +
\frac{2}{3} \, \frac{b^1_1}{a^2_1} \, (H_5 H'_5)^{-1} \right) } ,
\label{metric60}
\ee
that reduces to Eq.(\ref{EK11}) when $\theta_1=\pi/2$ and $r
\rightarrow 0$.  Now, using the transformations from the type IIB to
type IIA metrics we obtain the metric as seen by the string embedded
in the NS5-NS5'-D4 brane system
\be
ds^2 = -H_3^{-1/2} \, dt^2 + H_3^{1/2} \, H'_5  \, dx^2_4 ,
\ee
again, one can write the Nambu-Goto Lagrangian and derive
\bea
E &=& - \frac{\partial {\cal {L}}_{NG}}{\partial \kappa} = \kappa \,
\frac{1}{2 \pi \alpha'} \, \int_0^{2 \pi} d\sigma
\frac{H_3^{-1/2}}{\sqrt{H_3^{-1/2} \, \kappa^2 - \frac{1}{6} \, a^2_1
\, \omega^2_1 \, H_3^{1/2} \, H'_5 }} , \nonumber \\
J_1 &=& \frac{\partial {\cal {L}}_{NG}}{\partial \omega_1} =
\frac{\omega_1 \, a^2_1}{6} \, \frac{1}{2 \pi \alpha'} \, \int_0^{2
\pi} d\sigma \frac{H_3^{1/2} \, H'_5 }{\sqrt{H_3^{-1/2} \, \kappa^2 - \frac{1}{6} \,
a^2_1 \, \omega^2_1 \, H_3^{1/2} \, H'_5 }} . \nonumber \\
&&
\eea
Thus,
\be
E = \frac{6 \, \kappa \, J_1}{a^2_1 \, \omega_1 \, H_3 \, H'_5  } .\label{E-J22}
\ee
After using T and S-duality transformations we get the type IIB
D5-D5'-D3 system, obtaining again Eq.(\ref{E-J22}).  This equation
coincides with Eqs.(\ref{EK11}) and (\ref{metric60}) for $\theta_1
= \pi/2$ corresponding to the equatorial rotation in $S^2$ in $T^{1,1}$.

~

~

{\it A spinning closed string in a 2-cycle of $T^{1,1}$}

~

Now we consider a closed string whose center of mass is not moving on
$T^{1,1}$ nor in $AdS_5$, but the string is spinning around some
isolated point, and it is stretched in $\theta_1$ (of $S^2$
in the base of the conifold). Without loss of generality, let us
assume that the center of the string is at the North pole.  A
convenient ansatz is
\bea
&& t = e \, \tau, \,\,\,\,\,\,\,\, \phi_1= e \, \omega_1 \, \tau,
\,\,\,\,\,\,\,\, \theta_1 = \theta_1(\sigma), \,\,\,\,\,\,\,\,
\rho(\sigma)=\rho(\sigma+\pi), \nonumber
\\ && \phi_2 = constant , \,\,\,\,\,\,\,\, \theta_2 = constant ,
\,\,\,\,\,\,\,\, \beta_i = 0 \,\,\,\,\, i=1, 2, 3 .
\eea
The corresponding metric (for the embedding of our string) reads
\be
ds^2 = L^2 [- e^2 \, d\tau^2 + \frac{1}{6} \, (d\theta_1^2 +
 \sin^2\theta_1 \, d\phi_1^2) + \frac{1}{9} \, \cos^2\theta_1 \, d\phi_1^2] .
\ee
Using the above ansatz on the second conformal constraint, we get
\be
- L^2 \, e^2 + \frac{L^2}{6} \, ({\theta_{1,}}_{\sigma}^2 +
 \sin^2\theta_1 \, e^2 \, \omega_1^2) + \frac{L^2}{9} \,
 \cos^2\theta_1 \, e^2 \, \omega_1^2 = 0 .
\ee
Therefore, we obtain the relation
\be
\frac{{\theta_{1,}}_{\sigma}^2}{6} = e^2 [1 - (\frac{1}{18} \,
 \sin^2\theta_1  + \frac{1}{9}) \, \omega_1^2] ,
\ee
while the Nambu-Goto Lagrangian now reads
\be
{\cal {L}}_{NG}=- \frac{\sqrt{\lambda}}{2 \pi} \,
\frac{e}{\sqrt{6}} \, \int_0^{2 \pi} d\sigma \,
{\theta_{1,}}_{\sigma} \, \sqrt{1 - \frac{\omega_1^2}{18} \, (2 +
\sin^2\theta_1)} \, ,
\ee
where $\sqrt{\lambda}=L^2/\alpha'$. We can calculate the momentum
conjugate to $\theta_1$
\be
P_1 = \frac{\partial {\cal {L}}_{NG}}{\partial \theta_{1,\sigma}}
= - \frac{\sqrt{\lambda}}{2 \pi} \, \frac{e}{\sqrt{6}} \,
\int_0^{2 \pi} \, d\sigma \, \sqrt{1-\frac{\omega_1^2}{18} \,
(2+\sin^2\theta_1)} ,
\ee
thus, for the maximum $\theta_1=\pi/2$, $dP_1/d\sigma=0$, it
corresponds to $\omega^2_1=6$.  The energy and R-symmetry charge
are given by
\bea
E &=& \frac{2 \, \sqrt{\lambda}}{\pi} \, \frac{e}{\sqrt{6}} \,
\int_0^{\theta_0} \,
\frac{d\theta_1}{\sqrt{1-\frac{\omega^2_1}{18}(2+\sin^2\theta_1)}}
, \nonumber \\
J_1 &=& \frac{2 \, \sqrt{\lambda}}{\pi} \, \frac{\omega_1}{18} \,
\frac{e}{\sqrt{6}} \, \int_0^{\theta_0} \, d\theta_1 \,
\frac{(2+\sin^2\theta_1)}{\sqrt{1-\frac{\omega^2_1}{18}(2+\sin^2\theta_1)}}
,
\eea
using the maximum value for $\theta_1$ one can easily get
\be
\frac{E}{\sqrt{6}} = J_1 + \frac{e}{3 \, \sqrt{3}} \,
\frac{\sqrt{\lambda}}{\pi} . \label{E-J23}
\ee
Now, we consider how this relation transforms in the intersecting
branes setup. After using the definitions
\be
dx^4 \rightarrow \frac{e \, \omega_1}{\sqrt{6}} \, \sin\theta_1 \,
d\tau , \,\,\,\,\,\,\,\,\,\,\, dx^5 \rightarrow \frac{1}{\sqrt{6}} \,
d\theta_1 , \,\,\,\,\,\,\,\,\,\,\, B_{64} = \sqrt{\frac{2}{3}} \,
\cot\theta_1 ,
\ee
we obtain a type IIB metric with non-trivial $B$-fields turned on.
Since, the $B$-fields depend on $\sigma$ through $\theta_1$, one
would expect they lead to two kind of contributions to the energy,
also modifying the relation $E-J$. One of the contributions should
come from the $B$-field term in the string action, however this is
proportional to
\be
B_{64} \, \partial_a x^6 \, \partial_b x^4 ,
\ee
which is trivially zero, since in the parametrization above $x^6$ has
been taken to be a constant. In order for the string to be coupled to
that $B$-field, it is necessary that its center of mass be
boosted along $x^6$, {\it i.e.} string spins in $S^2$ and moves
along the fiber $\psi$ as well. The actual contribution to energy and
$J_1$ comes directly from the type IIB metric (\ref{metricIIBnew})
\bea
ds^2 & =& - e^2 \, (H_3)^{-1/2} \, d\tau^2 \nonumber \\
&& + (H_3)^{1/2} \, [\frac{1}{6}
\, H'_5 \, (e^2 \, \omega^2_1 \, \sin^2\theta_1 \, d\tau^2 +
d\theta^2_1) + \frac{1}{9} \, e^2 \, \omega^2_1 \, (H_5 \, H'_5)^{-1}
\, \cos^2\theta_1 \, d\tau^2] ,
\eea
thus, we have
\bea
E &=& \frac{e}{2 \pi \alpha'} \, \int_0^{2 \pi} \,
\frac{d\sigma}{\sqrt{6}} \, \frac{(H_3)^{-1/4} \, (H'_5)^{1/2} \,
\theta_{1, \sigma}}{\sqrt{(H_3)^{-1/2} - (H_3)^{1/2} \, [\frac{1}{6}
\, H'_5 \, \sin^2\theta_1 + \frac{1}{9} \, (H_5 H'_5)^{-1} \,
\cos^2\theta_1] \, \omega^2_1}} \nonumber \\
& & \nonumber \\
J_1 &=& \frac{\omega_1 \, e}{2 \pi \alpha'} \, \int_0^{2 \pi} \,
\frac{d\sigma}{\sqrt{6}} \, \frac{ (H_3)^{3/4} \, (H'_5)^{1/2} \,
\theta_{1, \sigma} \, [\frac{1}{6} \, H'_5 \, \sin^2\theta_1 +
\frac{1}{9} \, (H_5 H'_5)^{-1} \,
\cos^2\theta_1]}{\sqrt{(H_3)^{-1/2} - (H_3)^{1/2} \, [\frac{1}{6}
\, H'_5 \, \sin^2\theta_1 + \frac{1}{9} \, (H_5 H'_5)^{-1} \,
\cos^2\theta_1] \, \omega^2_1}}.
\eea
In the limit when $x^7 \rightarrow 0$ the relation (\ref{E-J23})
is recovered.  The corresponding type IIA metric is obtained through
the relation between type IIA and type IIB metrics, and it corresponds
to the NS5-NS5'-D4 intersecting brane system. The embedding of this
spinning closed string solution in this metric is
\be
ds^2  = - e^2 \, (H_3)^{-1/2} \, d\tau^2
+ H'_5 (H_3)^{1/2} \, [\frac{1}{6}
\, (e^2 \, \omega^2_1 \, \sin^2\theta_1 \, d\tau^2 +
d\theta^2_1)] , \label{metric77}
\ee
which corresponds to a closed string boosted along $x^4$. After T
and S-duality transformations the metric as seen by the spinning
string in the D5-D5'-D3 configuration turns out to be
\be
ds^2 = - e^2 \, (H_3 H_5 H'_5)^{-1/2} \, d\tau^2
+ (H'_5 H_3)^{1/2} \, (H_5)^{-1/2} \, [\frac{1}{6}
\, (e^2 \, \omega^2_1 \, \sin^2\theta_1 \, d\tau^2 +
d\theta^2_1)] , \label{metric78}
\ee
also corresponding to a closed string boosted parallel to $(x^4,
x^5)$ plane.  For $\theta_1=\pi/2$ and $x^7 \rightarrow 0$, the
$E-J$ relations derived from metric (\ref{metric77}) and
(\ref{metric78}) become Eq.(\ref{E-J23}).

\subsection{Spinning closed strings in $AdS_5$ and boosted along $\psi$ in $T^{1,1}$}

Here we consider the situation where a closed string spins centered in
$S^3$ parametrizing $AdS_5$, and its center of mass moves along $\psi$
in $T^{1,1}$. This is the most interesting situation in terms of
analyzing IR and UV limits in the dual ${\cal {N}}=1$ SYM theory.
For the embedding in the $AdS_5 \times T^{1,1}$ metric we use the
ansatz
\bea
&& t = \kappa \, \tau , \,\,\,\,\,\,\,\, \beta_3 \equiv \phi= \omega
\, \tau , \,\,\,\,\,\,\,\, \psi \equiv \varphi = \nu \, \tau ,
\,\,\,\,\,\,\,\, \rho(\sigma)\equiv\rho(\sigma+2 \pi) \nonumber
\\ && \phi_i = constant , \,\,\,\,\,\,\,\, \theta_i = constant ,
\,\,\,\,\,\,\,\, \beta_i = 0 \,\,\,\,\, i=1, 2 . \label{ans1}
\eea
We start from the conifold metric (\ref{conifoldmetricglobal}) and using
conformal constraints it leads to
\be
\left( \frac{d\rho}{d\sigma} \right)^2 = \kappa^2 \, \cosh^2\rho -
\omega^2 \, \sinh^2\rho - \frac{\nu^2}{9} . \label{eq80}
\ee
The second derivative is exactly given by
Eq.(\ref{secondderivative}). Notice that for multi-spin string
solutions even a more general rotating string ansatz is possible,
if for instance we consider the center of mass of the string
boosted along $\theta_i = \eta_i \, \tau$ and $\phi_i = \Omega_i
\, \tau$ for $i=1, 2$. Now, for simplicity we take
$\theta_i=\pi/2$, then $\nu^2/9 \rightarrow \nu^2/9 +
(\Omega_1^2+\Omega_2^2)/6$, and therefore we can proceed in a
similar way as using the ansatz (\ref{ans1}), but considering
several angular momenta.  This problem resembles the situations
discussed in \cite{Frolov:2003qc} and it deserves further study in
the context of finding new non-BPS sectors in the dual field
theory. The integral of Eq.(\ref{eq80}) leads to
\be
2 \, \pi = \int_0^{2 \pi} \, d\sigma = 4 \, \int_0^{\rho_0} \, d\rho
\, \frac{1}{\sqrt{(\kappa^2-\frac{\nu^2}{9}) \, \cosh^2\rho -
(\omega^2 - \frac{\nu^2}{9}) \, \sinh^2\rho}}.
\ee
In analogy with the case of $AdS_5 \times S^5$ there are three
conserved quantities, energy, spin and R-symmetry charge $J$,
\bea
E&=&\sqrt{\lambda} \, \kappa \, \int_0^{2\pi} \frac{d\sigma}{2 \pi} \,
\cosh^2\rho \equiv \sqrt{\lambda} \, {\cal {E}},  \nonumber \\
S&=&\sqrt{\lambda} \, \omega \, \int_0^{2\pi} \frac{d\sigma}{2 \pi} \,
\sinh^2\rho \equiv \sqrt{\lambda} \, {\cal {S}}, \nonumber \\
J&=&\sqrt{\lambda} \, \frac{\nu}{9} \, \int_0^{2\pi} \frac{d\sigma}{2
\pi} \equiv \sqrt{\lambda} \, \frac{\nu}{9}, \label{int82}
\eea
thus, we have the relation
\be
E=9 \, \frac{\kappa}{\nu} \, J + \frac{\kappa}{\omega} \, S. \label{EJS}
\ee
Now, one can explicitly calculate the above integrals in terms of
hyper-geometric functions. It is useful to write down the energy-spin
relations for both, short and long strings.  We introduce the
parameter $\eta > 0$
\be
\coth^2 \rho_0 = \frac{\omega^2 - \frac{\nu^2}{9}}{\kappa^2 -
\frac{\nu^2}{9}} = 1 + \eta , \label{def-eta}
\ee
where $\rho_0$ is the maximum value of $\rho$ for a single folded string.
Therefore, from Eqs.(\ref{int82}) we obtain the following relations
\bea
\left(\kappa^2 - \frac{\nu^2}{9}\right)^{1/2} & = & \frac{1}{\sqrt{\eta}} \,
_2F_1\left( \frac{1}{2}, \frac{1}{2}; 1;  -\frac{1}{\eta}  \right) , \\
{\cal {E}} & = & \frac{\kappa}{\sqrt{\eta} \, \sqrt{\kappa^2 -
\frac{\nu^2}{9}}} \, _2F_1\left( -\frac{1}{2}, \frac{1}{2}; 1;
-\frac{1}{\eta} \right) , \\
{\cal {S}} & = & \frac{\omega}{2 \, \eta \, \sqrt{\eta} \,
\sqrt{\kappa^2 - \frac{\nu^2}{9}}} \, _2F_1\left( \frac{1}{2},
\frac{3}{2}; 2; -\frac{1}{\eta} \right) .
\eea
Now, we analyze the short and long string limits.

~

~

{\it Short strings}

~

For $\rho_0 \rightarrow 0$ we have $\eta >> 1$ from above
expressions and the definition of $\eta$ we get the relations
\be
\kappa^2 \approx \frac{\nu^2}{9} + \frac{1}{\eta} ,
\,\,\,\,\,\,\,\, \omega^2 \approx \kappa^2 + 1 \approx
\frac{\nu^2}{9} + 1 + \frac{1}{\eta} ,
\ee
thus $\omega^2 - \kappa^2 \approx 1$, and using the expression for
$\rho'$, an approximate short string solution satisfies
\be
\sinh^2\rho \approx \rho^2 \approx \rho_0^2 \, \sin^2\sigma \approx
\frac{1}{\eta} \, \sin^2\sigma ,
\ee
therefore
\be
\frac{1}{\eta} \approx \frac{2 {\cal {S}}}{\sqrt{1 + \frac{\nu^2}{9}}}
<< 1 .
\ee
Then, from Eq.(\ref{EJS}) it is obtained a relation similar to
Eq.(\ref{ES})
\be
{\cal {E}} \approx 9 \, \sqrt{\frac{\nu^2}{9} + \frac{2 \, {\cal
{S}}}{\sqrt{1+\frac{\nu^2}{9}}}} + \frac{\sqrt{\frac{\nu^2}{9} +
\frac{2 \, {\cal {S}}}{\sqrt{1+\frac{\nu^2}{9}}}}} {\sqrt{1 +
\frac{\nu^2}{9} + \frac{2 \, {\cal {S}}}{\sqrt{1+\frac{\nu^2}{9}}}}}
\,\, {\cal {S}} ,
\label{ES1}
\ee
For $\nu << 1$, then ${\cal {S}} << 1$ and Eq.(\ref{ES1}) reduces to
\be
E^2 \approx 9 \, J^2 + 81 \, (2 \, \sqrt{\lambda}) \, S.
\label{es12}
\ee
All these expressions become the corresponding ones of
spinning strings in $S^3$ of $AdS_5$ in $AdS_5 \times S^5$ when we
take $\nu^2/9 \rightarrow \nu^2$. However, we have to remember that
$\sqrt{\lambda}$ corresponds now to $L^2/\alpha'$, where $L^2 =
\sqrt{27/16} R^2$.  This shows that the scale is set by the conifold
instead of $S^5$.  Eq.(\ref{es12}) is the limit for short strings in
$AdS_5 \times T^{1,1}$. They probe the small curvature region of
$AdS_5$. Moreover, if the boost energy is much smaller than the
rotational energy, {\it i.e.} $\nu^2<< 6 \, {\cal {S}}$, then
\be
{\cal {E}} \approx 9 \, \sqrt{2 \, {\cal {S}}} +
\frac{\nu^2}{2 \, \sqrt{ 2 \, {\cal {S}}}} ,
\ee
which after scaling $\nu^2/9 \rightarrow \nu^2$ is the flat-space
Regge trajectory, as in the case of $AdS_5 \times S^5$. On the other
hand, when the boost energy is greater than the spin ($6 {\cal
{S}}<<\nu$)
\be
E \approx 3 \, J + S + 3 \, \left( 27 - \frac{1}{2} \right) \,
\frac{\lambda \, S}{J^2}. \label{classicalen}
\ee
This corresponds to a short string spinning slowly around $S^3$ in
$AdS_5$ and boosted very fast along $\psi$. It can be related to
the leading quantum term in the spectrum of $AdS_5 \times
T^{1,1}$, in the frame boosted to the speed of light along the
fiber $\psi$. As in the case of $AdS_5 \times S^5$, the classical
energy (\ref{classicalen}) of the spinning string boosted along
$\psi$ should also be obtained from the quantum spectrum of string
oscillators in that boosted frame. Now, if we are in a sector
where $1 << S << J$, the quantum spectrum reproduces the classical
energy. Specifically, the $S$ term is related to the mass term in
light-cone string coordinates, in the plane wave background. It is
related to the curvature of the anti de Sitter space.

~

~

{\it Long strings}

~

For long strings $\rho_0$ is large, what implies that
$\eta << 1$ and therefore,
\bea
\kappa^2 & \approx & \frac{\nu^2}{9} + \frac{1}{\pi^2} \, \log^2(1/\eta) , \\
\omega^2 & \approx & \frac{\nu^2}{9} + \frac{1}{\pi^2} \, (1+\eta)
\, \log^2(1/\eta) , \\
\eea
and
\be
{\cal {S}} \approx \frac{2 \omega}{\eta \, \log\frac{1}{\eta}} .
\ee
Since, as in the case of $AdS_5 \times S^5$ there is no simple
relation between energy and spin, we must consider two limits. If $\nu
<< 3 \, \log (1/\eta)$, it is obtained the relation
\be
E \approx S + \frac{9 \, \sqrt{\lambda}}{\pi} \, \log (S/\sqrt{\lambda}) +
\frac{\pi J^2}{2 \sqrt{\lambda} \log (S/\sqrt{\lambda})} ,
\ee
while if $\log ( 3\, {\cal {S}}/\nu) << 3 \, \nu << {\cal {S}}$, then
\be
E \approx 3 \, J + S + \frac{27 \, \lambda}{2 \, \pi^2 \, J} \, \log^2 (S/J) .
\ee

~

~

{\it Spinning strings and brane constructions}

~

Let us consider the following ansatz for a closed string spinning in
the $(x_1, x_2)$ plane and boosted along the direction $x_6$, embedded
in the metric (\ref{metricIIBnew}) as follows
\be
x_0 = \kappa \, \tau , \,\,\,\,\,\,\, \gamma_3 = \omega \, \tau , \,\,\,\,\,\,\,
x_6 = \frac{\nu}{3} \, \tau , \,\,\,\,\,\,\,\, x^7(\sigma) = x^7(\sigma + 2 \pi) ,
\label{an1}
\ee
where $\gamma_3$ parametrizes the rotation in $(x_1,x_2)$, and the
rest of coordinates are fixed. Also, we consider the string is
stretched along $x^7$.

The type IIB metric (\ref{metricIIBnew}) as seen by the string is given by
\be
ds^2 = (H_3)^{-1/2} \, (- \kappa^2 + \omega^2) \, d\tau^2 +
(H_3)^{1/2} \, (H_5 \, H'_5)^{-1} \, \frac{\nu^2}{9} \, d\tau^2
+ H_3^{1/2} \, (H_5 \, H'_5) \, dx_7^2 .
\ee
There is no contributions from $B$-fields since $x_4$ and $x_8$ are
constants. Since we are interested in the type IIA NS5-NS5'-D4 brane
construction, we write the corresponding type IIA metric as follows
\be
ds^2 = (H_3)^{-1/2} \, (- \kappa^2 + \omega^2) \, d\tau^2 +
(H_3)^{-1/2} \, (H_5 \, H'_5) \, \frac{\nu^2}{9} \, d\tau^2
+ H_3^{1/2} \, (H_5 \, H'_5) \, dx_7^2 .
\ee
>From the second conformal constraint it is obtained
\be
\left(\frac{dx_7(\sigma)}{d\sigma}\right)^2 = \frac{1}{H_3 \, H_5 \,
H'_5} \, \left(\kappa^2 - \omega^2 - H_5 \, H'_5 \, \frac{\nu^2}{9} \right),
\ee
and therefore
\be
2 \, \pi = \int_0^{2 \pi} d\sigma = 4 \int_0^{x^7_0} \frac{dx_7 \,
\sqrt{H_3 \, H_5 \, H'_5}}{ \sqrt{\kappa^2 - \omega^2 - H_5 \, H'_5 \,
\frac{\nu^2}{9}}} .
\ee
We can also derive explicit expressions for the energy, spin
and R-symmetry charge
\bea
E &=& 4 \, \frac{\kappa}{2 \pi \alpha'} \, \int_0^{x_0^7} \frac{dx_7 \,
\sqrt{H_5 \, H'_5}}{ \sqrt{\kappa^2 - \omega^2 - H_5 \, H'_5 \,
\frac{\nu^2}{9}}} , \nonumber \\
J &=& 4 \, \frac{\nu}{18 \pi \alpha'} \, \int_0^{x_0^7} \frac{dx_7 \,
(H_5 \, H'_5)^{3/2}}{ \sqrt{\kappa^2 - \omega^2 - H_5 \, H'_5 \,
\frac{\nu^2}{9}}} , \nonumber \\
S &=& 4 \, \frac{\omega}{2 \pi \alpha'} \, \int_0^{x_0^7} \frac{dx_7 \,
\sqrt{H_5 \, H'_5}}{ \sqrt{\kappa^2 - \omega^2 - H_5 \, H'_5 \,
\frac{\nu^2}{9}}} .
\eea
In all the expressions above $x^7_0$ is the maximum length of a single folded
string.

~

~

In addition, we can consider another kind of closed string
spinning in the $(x_1, x_2)$ plane, boosted along the direction
$x_6$.  In this case we consider the rotation of the string in the
$(x_1, x_2)$ plane parametrized in polar coordinates $(r, \phi)$
and also stretched along $r$ in this plane, but unlike the example
above, this is not stretched along $x_7$.
\be
x_0 = \kappa \, \tau , \,\,\,\,\,\,\, \phi = \omega \, \tau , \,\,\,\,\,\,\,
x_6 = \frac{\nu}{3} \, \tau , \,\,\,\,\,\,\,\, r(\sigma) = r(\sigma + 2 \pi) .
\label{an2}
\ee
The metric of the type IIA NS5-NS5'-D4 brane construction, as seen by
the string, becomes
\be
ds^2 = (H_3)^{-1/2} \, [(- \kappa^2 + \omega^2 \, r^2(\sigma)) \, d\tau^2 +
r'^2 \, d\sigma^2] +
(H_3)^{-1/2} \, (H_5 \, H'_5) \, \frac{\nu^2}{9} \, d\tau^2 .
\ee
>From the second conformal constraint we get
\be
\left(\frac{dr(\sigma)}{d\sigma}\right)^2 = \kappa^2 - H_5 \, H'_5
\, \frac{\nu^2}{9} - \omega^2 \, r^2(\sigma)  ,
\ee
and
\be
2 \, \pi = \int_0^{2 \pi} d\sigma = 4 \int_0^{r_0} \frac{dr}{
\sqrt{\kappa^2 - H_5 \, H'_5 \, \frac{\nu^2}{9} - \omega^2 \,
r^2(\sigma)}} .
\ee
The expressions for the energy, spin and R-symmetry charge are
\bea
E &=& 4 \, \frac{\kappa \, (H_3)^{-1/2}}{2 \pi \alpha'} \,
\int_0^{r_0} \frac{dr}{ \sqrt{\kappa^2 - H_5 \, H'_5 \,
\frac{\nu^2}{9} - \omega^2 \, r^2(\sigma)}} , \nonumber \\
J &=& 4 \, \frac{\nu \, (H_3)^{-1/2}}{18 \pi \alpha'} \,
\int_0^{r_0} \frac{dr \, \, (H_5 \, H'_5)}{ \sqrt{\kappa^2 - H_5
\, H'_5 \, \frac{\nu^2}{9} - \omega^2 \, r^2(\sigma)}} , \nonumber
\\
S &=& 4 \, \frac{\omega \, (H_3)^{-1/2}}{2 \pi \alpha'} \,
\int_0^{r_0} \frac{dr \,\, r^2}{ \sqrt{\kappa^2 - H_5 \, H'_5 \,
\frac{\nu^2}{9} - \omega^2 \, r^2(\sigma)}} .
\eea

For these particular string solutions we have not succeeded in
obtaining explicit energy-spin and energy-charge relations as for the
previous cases. Probably, the reason could be that our ansatze
(\ref{an1}) and (\ref{an2}) for the string solutions embedded in the
intersecting brane backgrounds, do not properly capture the
corresponding features for large spin and large R-symmetry charge
operators in the dual SYM theory.  It would be very interesting to
explore more general string solutions embedded in the above brane
constructions related to the conifold background, and identifying
those solitons with the corresponding dual SYM theory operators.


\section{Classical string solutions and ${\cal N}=1$ SYM}

We start with a review of the Klebanov-Tseytlin solution
\cite{Klebanov:2000nc}. This takes into account the back-reaction due
to the presence of $M$ fractional D3-branes in the set up given by $N$
regular D3-branes placed at the conifold singularity, {\it i.e.} it
includes higher order corrections in $M/N$. Starting from the type IIB
supergravity action (\ref{actionIIB}) (see Appendix A), it has been
proposed the following ansatz for the Einstein frame metric in 10
dimension
\bea
ds^2 &=& L^2 \, \{e^{-\frac{2}{3} (B+4C)} \, (du^2 + e^{2 A(u)}
dx_m \, dx^m) \nonumber \\
&+& \frac{1}{9} \, e^{2B} \, (d\psi + \sum^2_{i=1} \cos\theta_i
d\phi_i)^2 + \frac{1}{6} \, e^{2 C} \, \sum^2_{i=1} (d\theta^2_i +
\sin^2 \theta_i d\phi_i^2))\} \, , \label{metrict11}
\eea
where $B=B(u)$ and $C=C(u)$, in the limit when these functions
vanish the metric of $AdS_5 \times T^{1,1}$ is recovered. The
radius $L$ is proportional to $N^{1/4}$. The case studied in
\cite{Klebanov:2000nc} assumes the axion ${\cal {C}}=0$, that
trivially makes zero the kinetic terms for the axion in
Eqs.(\ref{eom-einstein-IIb}) and (\ref{eom-dilaton-IIb}), and it
implies that $* F_3 \, \wedge \, H_3=0$. On the other hand, a RR 3
form flux through the 3-cycle in $T^{1,1}$ is created due to the
presence of the fractional branes \cite{Klebanov:1999rd}. This is
$F_3=d{\cal C}_2$, and it turns out to be proportional to the
closed 3-form constructed in \cite{Gubser:1998fp}. Therefore, we
have
\be
F_3 = P \, e^\psi \, \wedge \, \omega_2 \, , \label{f3}
\ee
where $P$ is a constant proportional to $M$, and
\be
\omega_2 \equiv \frac{1}{\sqrt{2}} \, (e^{\theta_1} \, \wedge
e^{\phi_1} - e^{\theta_2} \, \wedge e^{\phi_2}) \, .
\ee
Particularly, in the normalization used here where $L=1$, $P$
results to be proportional to $M/N$. We also use the orthonormal
basis of 1-forms of \cite{Gubser:1998fp}. The NS-NS 2-form
potential is proportional to the 2-form
\bea
{\cal B}_2 &=& T(u) \, \omega_2 \, , \\
H_3 &=& T'(u) du \, \wedge \, \omega_2 \, . \label{h3}
\eea
In addition, the 5-form field strength can be written as $F_5={\cal
F} + * {\cal F}$, where ${\cal F}=K(u) e^\psi \wedge e^{\theta_1}
\wedge e^{\phi_1} \wedge e^{\theta_2} \wedge e^{\phi_2}$, while $*
{\cal F}= e^{4A-(8/3)(B+4C)} \, K(u) du \wedge dx^1 \wedge dx^2
\wedge dx^3 \wedge dx^4$. Therefore, using the ansatz (\ref{f3})
for $F_3$, Eq.(\ref{eom-f3}) is satisfied since $F_5 \wedge
H_3=0$.

Eq.(\ref{eom-f5}) gives a relation between the functions $T(u)$
and $K(u)$, $dK/du=P \, dT/du$, and therefore the integral $K(u)=P
\, T(u) + Q$.

>From Eq.(\ref{eom-h3}) it is obtained $\nabla^2 \phi=1/12 \,
(e^\phi \, F_3^2 - e^{-\phi} \, H_3^2)$, which defines the dilaton
dependence on the variable $u$.

As was explained in \cite{Klebanov:2000nc}, the equations above can
be derived from the following 5d action
\be
S_5 = - \frac{2}{\kappa^2} \, \int d^5x \, \sqrt{g_5} \, \left[
\frac{1}{4} \, R_5 - \frac{1}{2} \, G_{ab}(\varphi) \,
\partial\varphi^a \, \partial\varphi^b - V(\varphi) \right] \, ,
\label{S5d}
\ee
where
\bea
G_{ab}(\varphi) \, \partial\varphi^a \, \partial\varphi^b &=& 15
\, (\partial q)^2 + 10 \, (\partial f)^2 + \frac{1}{4} \,
(\partial\phi)^2 + \frac{1}{4} \, e^{-\phi-4 f -6q} \, (\partial
T)^2 \, , \\
V(\varphi) &=& e^{-8q} \, (e^{-12f}-6 \, e^{-2f}) + \frac{1}{8} \,
P^2 \, e^{\phi+4f-14q} + \frac{1}{8} \, (Q+PT)^2 \, e^{-20q} \, .
\eea
The scalar fields $\varphi^a=(q, f, \phi, T)$ include the
combinations
\be
q=\frac{2}{15} \, (B+4C) \, , \,\,\,\,\,\,\,\,\,\,\,\,\,\,\,\,\,
f=-\frac{1}{5} \, (B-C) \, ,
\ee
which measure the volume and the ratio of scales of the internal
manifold $T^{1,1}$.

The simplest fixed-point solution for $P=0$ corresponds to
$AdS_5$. On the other hand, the action (\ref{S5d}) generates a
system of second order differential equations. If one is
interested in preserving certain amount of supersymmetry the
second order differential equations can be replaced by a system of
first order ones
\bea
\varphi'^a &=& \frac{1}{2} G^{ab} \, \frac{\partial W}{\partial
\varphi^b} \, , \\
A' &=& - \frac{1}{3} \, W(\varphi) \, , \label{derivatives}
\eea
where $A(u)$ was defined in the metric (\ref{metrict11}), and the
superpotential $W$ is a function of the scalars and satisfies
\be
V = \frac{1}{8} \, G^{ab} \, \frac{\partial W}{\partial\varphi^a}
,\ \frac{\partial W}{\partial\varphi^b} - \frac{1}{3} W^2 \, .
\ee
Using Eqs.(\ref{derivatives}) one gets the following set of first
order equations
\bea
T' &=& P \, e^{4(f-q)} \, , \label{firstordersystem1} \\
f' &=& \frac{3}{5} \, e^{4(f-q)} \, (e^{-10 q} - 1) \, ,
\label{firstordersystem2} \\
q' &=& \frac{2}{15} \, e^{4(f-q)} \, (3 + 2 \, e{-10 f}) -
\frac{1}{6} \, (Q + P \, T) \, e^{-10 q} \, ,
\label{firstordersystem3}
\eea
by solving them it is possible to discuss several physical aspects
as follows.

As was mentioned, the simplest solution that one can obtain is a
fixed-point solution. The system
(\ref{firstordersystem1}-\ref{firstordersystem3}) is satisfied for any
constant $T$, and for $Q=4$ it straightforwardly implies that $q=f=0$,
and consequently $B=C=0$ in the metric (\ref{metrict11}), leading to
the $AdS_5 \times T^{1,1}$ background. In this case the solutions
discussed in section 3, as well as their related brane constructions
are valid.

A very important feature of the first order differential equations
system is the existence of general solutions depending on $u$, and
leading to an RG flow in the dual field theoretical
interpretation. Thus, let us assume that above certain UV cut-off
$u_0$ there exists a SCFT, {\it i.e.} the total number of
fractional branes, and therefore $P$, are zero. One can reach this
configuration by placing $M$ fractional anti D3-branes at $u_0$.
Certainly, it leads to vanishing values for the functions $q$ and
$f$, while $T$ is a constant, at $u=u_0$. Moreover, this setting
automatically generates a null function $f(u)=0$ after integration
of Eq.(\ref{firstordersystem2}), so that $B=C=0$ along the entire
RG flow toward the IR. This physically means that the internal
manifold $T^{1,1}$ is invariant during the whole RG flow, allowing
only to change its overall size.

With the above settings and introducing a new variable $Y(u)=e^{6
q(u)}$, the system
(\ref{firstordersystem1}-\ref{firstordersystem3}) becomes
\be
\frac{dY}{dK}=\frac{1}{P^2} \, (4 \, Y - K) \, ,
\ee
{\it i.e.} the derivative of the conformal factor related to the
$AdS_5$ space in the metric (\ref{metrict11}) with respect to the
$F_5$ strength. A general solution is
\be
Y= a_0 \, e^{4 K/P^2} +\frac{K}{4} +\frac{P^2}{16} \, ,
\label{Y}
\ee
where the constant $a_0$ has been chosen in order to satisfy the
UV boundary condition, $q=0 \Leftrightarrow Y=1$, and therefore
$K=K_0=Q+P T_0$ where $Q=4$,
\be
a_0 = -\left( \frac{P^2}{16} + \frac{P T_0}{4} \right) \,
e^{-16/P^2-4T_0/P} \, .
\ee
Essentially, one can choose the metric to approach the canonical
$AdS_5$ as $u\rightarrow u_0$ ($A(u)\rightarrow u$), leading to
\be
ds^2_{10}= e^{-5 q} \, du^2 + e^{2 A-5 q} \, dx_n dx_n + e^{3 q}
\, dS^2_{T^{1,1}} \, , \label{metricA}
\ee
where in general
\be
A(u)=A_0+q(u)+\frac{2}{3} f(u) + \frac{1}{P} \, T(u) \, .
\ee
As we have seen $f(u)=0$ along the RG flow.

Beyond the near horizon approximation it is possible to construct
asymptotically flat solutions of the system
(\ref{firstordersystem1}-\ref{firstordersystem3}). For $P=0$, {\it
i.e.} $M=0$, the solution describes regular D3-branes at the
conifold singularity. On the other hand, if $P\neq 0$ there is a
generalization with logarithmic running charge. Therefore, let us
consider the metric written in the form
\be
ds^2_{10} = s^{-1/2}(r) \, dx_n dx_n + h^{1/2}(r) \, (dr^2 + r^2
\, dS^2_{T^{1,1}}) \, . \label{metricB}
\ee
Comparing this metric with the one of Eq.(\ref{metricA}) one gets
\be
s(r) = e^{10 q(u) - 4 A(u)} \, , \,\,\,\,\,\,\,\,\,\, e^{6 q(u)} =
r^4 \, h(r) \, .
\ee
In addition
\be
\frac{dr}{r} = e^{-4 q(u)} \, du \, .
\ee
In order to obtain an explicit solution we start integrating $T'=
dT/du = P \, e^{-4 q}$, leading to
\be
T(r) = {\tilde T} + P \log r \, .
\label{T}
\ee
Now, using the expression for $Y$ given by Eq.(\ref{Y}) it becomes
\be
Y=e^{6q}=r \, h(r) = a_0 \, e^{4 {\tilde Q}/P^2 + 4 \log r} +
\frac{1}{4} ({\tilde Q} + P^2  \log r) + \frac{1}{16} \, P^2 \, .
\ee
>From this equation it can be obtained an explicit expression for
$h$
\be
h(r) = b_0 + \frac{k_0+P^2 \log r}{4 r^4} \, ,
\label{klebanov-tseytlin-h}
\ee
where $b_0=a_0 \, e^{4 {\tilde Q}/P^2}$ and $k_0={\tilde Q}+
P^2/4$. The solution for $s$ is $s(r)=h(r)$. Thus, the metric
(\ref{metricB}) becomes
\be
ds^2_{10} = h^{-1/2}(r) \, dx_n dx_n + h^{1/2}(r) \, (dr^2 + r^2
\, dS^2_{T^{1,1}}) \, , \label{metricBB}
\label{klebanov-tseytlin}
\ee
where
\bea
h(r) &=& b_0 + \frac{k_0+P^2 \log r}{4 r^4} \, , \\
k_0 &=& - P^2 \log r_* = Q + P {\tilde T} + \frac{1}{4} \, P^2 \,
.
\eea
Therefore, there is a logarithmic RG flow.

Now, let us consider how is the related type IIA brane
construction. First of all, we recall that in addition to the N
regular D3 branes, there are M fractional D3 branes located at
certain value of the radial coordinate at the conifold. On the
other hand, as we have seen at the UV a set of M-fractional anti
D3 branes, has been introduced at some radial distance $u_0$.
Thus, it effectively leads to an UV cut-off in the dual gauge
field theory, while at this UV cut-off the gauge theory is
conformal invariant. For this case, again, in the UV limit the
type IIB supergravity background is $AdS _5 \times T^{1,1}$. For
the $AdS _5 \times T^{1,1}$, one started with a configuration of
two perpendicular D5 branes and N D3 branes stretched between
them. Then, performing S-duality one gets two perpendicular NS5
fivebranes with N D3 branes stretched between them, and after
$T_3$ duality one gets the two perpendicular NS5 branes plus N D4
branes stretched between them. In the compact $x^6$ direction the
two NS5 branes are equally separated from both sides.  In the DM
construction the only overall direction is $x^7$, and this is
identified with the radial direction in the conifold supergravity
background, as $x^7=\log r$. In addition, one has the freedom to
include M fractional D4 branes, parallel to the regular ones but
separated in $x^7$ at a distance, let us say, $x^7_*$. Now this
leads to a bending of the perpendicular NS5 branes as a function
of $x^7$. Now, if we introduce M fractional anti-D4 branes at
certain $x^7_{UV}$, the bending of the NS5 branes will be
compensated beyond $x^7_{UV}$, and therefore recovering the
conifold.  At this UV point the results of section 3 are valid.

In addition, in ref.\cite{Tseytlin:2002ny} semi-classical solutions of
closed strings expanded around the solutions presented in section 2,
corresponding to near-conformal backgrounds, have been studied.

In order to explore closed string solutions around the above UV
fixed point one can define different ansatze, and use similar
calculations as we have done in section 3, in order to obtain
energy-spin and energy-R symmetry charge relations.  It would be
very interesting to extend these studies to the deep IR limit of
the Klebanov-Strassler solution. In this case spinning string
solutions has been recently studied \cite{Pons:2003ci}. Besides,
one very interesting question is whether solitonic string
solutions in the deep IR of Klebanov-Strassler background can be
mapped onto intersecting brane constructions. In principle, there
is not explicit supergravity solutions for these brane
constructions. This is in part due to the fact that the presence
of the M fractional D4 branes induce a bending on the NS5 branes
in the overall perpendicular direction.  A step forward in this
direction is to consider the case where no fractional D4 branes
are present. This case maps onto the deformed conifold
\cite{ohta}. We leave the investigation of classical string
solutions in these very interesting backgrounds and also to extend
the string theory states/SYM theory operators map for them for a
future work.


\section{Discussion}

We have studied classical solutions of closed strings boosted and
spinning in the conifold background, and also have obtained new
solitonic string solutions. After applying the Dasgupta-Mukhi map
we have analyzed how these solutions transform onto a background
of D4-branes stretched between perpendicular NS fivebranes in type
IIA string theory. In these cases, the configurations have certain
non-vanishing Neveu-Schwarz $B$-fields which modify the spectra
only in certain special cases, and particularly in the case of the
type IIB metric, in one of the steps of the DM construction. Then,
we have discussed about classical solutions of closed strings in
the UV limit of the Klebanov-Tseytlin background. By using a map
also inspired in the DM construction we have been able to
understand how the logarithmic running of the couplings in the
dual SYM theory is seen from the intersecting brane construction
viewpoint, and how classical closed string solutions can be
embedded in the UV limit of this background. We have also
commented on the possibility to extend these studies to the
Klebanov-Strassler solution, using as a previous step the
Ohta-Yokono construction, relating the deformed conifold and
intersecting brane constructions.

A very interesting issue that we have not addressed in this paper
concerns to the finding of the corresponding operators in the dual
${\cal {N}}=1$ SYM theories. Some comments are in order.
Following the proposal of Gubser, Klebanov and Polyakov, for the
case of the conifold solution one can firstly consider the case of
the dual conformal field theory to the conifold background
developed by Klebanov and Witten, {\it i.e.}  ${\cal {N}}=1$
$SU(N) \times SU(N)$ SYM theory, and consider twist two operators
${\cal {O}}_n$.  Those are the operators with the lowest conformal
dimension, and they are built out of $n$ covariant derivatives.
Classically, their conformal dimension is $\Delta_n = n+2$. Thus,
they are called twist 2. On the other hand, for these operators it
is expected that the leading term in the anomalous dimensions
receives a logarithmic correction to all orders in perturbation
theory and non-pertubatively as well
\cite{Korchemsky:1988si,Korchemsky:1992xv}. Therefore, $\Delta_n =
n + 2 + f(\lambda) \, \log n + \cdot \cdot \cdot $.  In the
expression above there is a functional dependence on the 't Hooft
coupling. In ref.\cite{Buchel:2002yq} it has been discussed how
the twist two operators should behave in the UV limit of the
Klebanov-Tseytlin solution for the compact $S^3$ theory
\cite{buchelS3}. However, the lack of a full non-linear solution
for the dual supergravity does not allow one to test the
logarithmic behavior.

On the other hand, after transforming the conifold onto the type
IIA NS5-NS5'-D4 intersecting brane system, a similar analysis
should be expected. In addition, these results should also be
valid in the case of the far UV limit of the Klebanov-Tseytlin
background. On the other hand, in terms of the analysis developed
by Buchel \cite{Buchel:2002yq} for the IR limit of the
Klebanov-Tseytlin background of \cite{buchelS3}, one could expect
analog results along the RG flow in Klebanov-Tseytlin case.
However, one must keep in mind that once the theory flows from the
conformal point, the identification above between the classical
rotating string energy and the conformal dimension is not
completely clear.

Besides, the NS5-NS5'-D4 brane configurations can be lifted to
M-theory descriptions \cite{Dasgupta:1998su,Dasgupta:1999wx}.
Essentially, in M-theory the NS5-NS5'-D4 system becomes two
perpendicular cylindrical M5 branes plus $N$ M5 branes, which are
compactified in the $x^{10}$ direction. It is very interesting to
consider a spinning and rotating closed M2 brane embedded in this
M-theory background. After the reduction to type IIA string
theory, the above system becomes a NS5-NS5'-D4 brane configuration
plus a spinning and rotating closed D2 brane embedded in the
corresponding background. Then, one can think of taking $T$
duality to obtain a type IIB NS5-NS5'-D3 brane configuration plus
a spinning and rotating closed D1 brane. After S-duality, one will
be left with D5-D5'-D3 system and a spinning closed F1 brane.
Therefore, we can see that other extended objects can also be
considered as solitonic solutions in the context of the Gubser,
Klebanov, Polyakov's proposal. It would be expected that these
solitonic solutions represent different SYM theory operators via
the proposed duality.  For instance, in the case of dual 3d SYM
theories, closed rotating M2 branes in $AdS_4 \times S^7$
\cite{Alishahiha:2002sy}, as well as, in $Mink_3 \times {\cal
{M}}$ \cite{Hartnoll:2002th}, where ${\cal {M}}$ is a special
holonomy manifold (for instance Spin(7), CY4 fold, Hyper-K\"ahler
manifolds \cite{Gursoy:2002tx}), have been recently considered.
One interesting remark is that for special holonomy manifolds,
classical rotating closed M2 branes lead to explicit classical
energy-spin and energy-charge relations that, if one trust the
correspondence beyond the conformal fixed-point, reproduce the
logarithmic behaviour of the anomalous dimension of twist two
operators in the dual SYM theories. Although, as we mentioned this
is related to dual 3d SYM theories, we would expect that a similar
behavior can be found for the M-theory descriptions related to the
type IIA NS5-NS5'-D4 brane systems that we study in this paper,
which are dual to 4d SYM theories.


~

~

\centerline{\large{\bf Acknowledgments}}

~

This work is dedicated to the memory of Ian I. Kogan.  We would
like to thank Harvey Meyer and Michael Teper for enlightening
discussions, and Diego Correa, Keshav Dasgupta, Sebastian Franco
and Kazutoshi Ohta for insightful comments. This work has been
supported by the PPARC Grant PPA/G/O/2000/00469, the Fundaci\'on
Antorchas, and the Consejo Nacional de Investigaciones
Cient\'{\i}ficas y T\'ecnicas (CO\-NI\-CET).

~
~

\subsection*{Appendix A: Type IIB supergravity in ten dimensions}

In order to define the notation in section 4 we briefly review the
action and equations of motion corresponding to ten dimensional
type IIB supergravity \cite{Schwarz:qr}. The field content of this
theory is given by the metric, a 4-form potential ${\cal C}_4$, a
scalar $\phi$, an axion ${\cal C}$, a R-R 2-form potential ${\cal
C}_2$, a NS-NS 2-form potential ${\cal B}_2$, two gravitinos with
the same chirality $\Psi^i_M$, and two dilatinos $\lambda_i$
($i=1,2$). Since there is not a simple covariant Lagrangian for
type IIB supergravity under the condition $F_5=*F_5$, one can
write a Lagrangian without constraining the five-form field
strength and, after derivation of the equations of motion, one can
impose that condition \cite{Bergshoeff:1995as}. We use the
notation given in \cite{Klebanov:2000nc}. Therefore, we consider
the bosonic type IIB supergravity action written in the Einstein
frame
\bea
S_{IIB} &=& -\frac{1}{2 \, \kappa^2_{10}} \int d^{10}x \,
\sqrt{-g_E} \, [ \, R_E - \frac{1}{2} (\partial\phi)^2 -
\frac{1}{12} \, e^{-\phi} \, (\partial{\cal B}_2)^2 - \frac{1}{2}
\, e^{2 \phi} \, (\partial{\cal C})^2 \nonumber \\
& & - \frac{1}{12} \, e^\phi \, (\partial{\cal C}_2-{\cal C} \,
\partial{\cal B}_2)^2 - \frac{1}{4 \cdot 5!} \, F^2_5 \, ] -
\frac{1}{2 \cdot 4! \cdot (3!)^2} \, \epsilon_{10} \, {\cal C}_4
\, \partial{\cal C}_2 \, \partial{\cal B}_2 \, ,
\label{actionIIB}
\eea
where
\bea
(\partial{\cal B}_2)_{MNK} &\equiv& 3 \,
\partial_{[M} \, {\cal B}_{NK]} \, , \nonumber \\
(\partial{\cal C}_4)_{MNKLP} &\equiv& 5 \,
\partial_{[M} \, {\cal C}_{NKLP]} \, , \nonumber \\
F_5 &\equiv& \partial{\cal C}_4 + 5 \, ({\cal B}_2
 \, \partial{\cal C}_2- {\cal C}_2 \, \partial{\cal B}_2) \, ,
\eea
plus the self-duality condition for $F_5$. The action of
Eq.(\ref{actionIIB}) can be rewritten using a Lagrangian given in
terms of differential forms, which leads to more compact
expressions of equations of motion. In the Einstein frame it reads
\cite{Tran:jj}
\bea
{\cal L}_{IIB} &=& R_E \, * {\bf {1}}- \frac{1}{2} \, * d\phi \,
\wedge \, d\phi - \frac{1}{2} \, e^{-\phi} \, * H_3 \, \wedge \,
H_3 - \frac{1}{2} \, e^{2 \phi} \, * d{\cal C} \, \wedge d{\cal C}
\, \nonumber
\\
& & - \frac{1}{2} \, e^\phi \, * F_3 \, \wedge \, F_3 -
\frac{1}{4} \, *  F_5 \, \wedge \, F_5  - \frac{1}{2} \, {\cal
C}_4 \, \wedge \, d{\cal C}_2 \, \wedge \, d{\cal B}_2 \, ,
\label{lagrangianIIB}
\eea
where we also have some definitions as follows
\bea
F_3 &\equiv& d{\cal C}_2 - {\cal C} \, d{\cal B}_2 \, , \nonumber
\\
H_3 &\equiv& d{\cal B}_2 \, , \nonumber \\
F_5 &\equiv& d{\cal C}_4 - \frac{1}{2} {\cal C}_2 \wedge d{\cal
B}_2 + \frac{1}{2} {\cal B}_2 \wedge d{\cal C}_2 \, .
\eea
The equations of motion are
\bea
R_{MN}&=&\frac{1}{2} \, \partial_M \phi \, \partial_N \phi +
\frac{1}{2} \, e^{2 \phi} \, \partial_M {\cal C} \, \partial_N
{\cal C} + \frac{1}{96} \, F_{MPQRS} \, F_N^{\,\,\,PQRS} \nonumber
\\
&& + \frac{1}{4} \, e^\phi \, (F_{MRS} \, F_N^{\,\,\,RS} -
\frac{1}{12} \, F_{RST} \, F^{RST} \, g_{MN}) \nonumber \\
&& + \frac{1}{4} \, e^{-\phi} \, (H_{MRS} \, H_N^{\,\,\,RS} -
\frac{1}{12} \, H_{RST} \, H^{RST} \, g_{MN}) \, ,
\label{eom-einstein-IIb} \\
d * d\phi &=& -  e^{2 \phi} \, * d{\cal C} \, \wedge \, d{\cal C}
- \frac{e^{\phi}}{2} \, * F_3 \, \wedge \, F_3 +
\frac{e^{-\phi}}{2} \, * H_3 \, \wedge \, H_3  \, ,
\label{eom-dilaton-IIb}
\\
d(e^{2 \phi} \, * d{\cal C}) &=& e^\phi \, * F_3 \, \wedge \, H_3
\, , \label{eom-axion-IIb} \\
d(e^{\phi} \, * F_3) &=& F_5 \, \wedge \, H_3 \, , \label{eom-f3}
\\
d(e^{- \phi} \, * H_3 - e^\phi \, {\cal C} \, * F_3) &=& - F_5 \,
\wedge \, (F_3 +  {\cal C} \, H_3) \, ,  \label{eom-h3} \\
d(* F_5) &=& - F_3 \, \wedge \, H_3 \, . \label{eom-f5}
\eea

~


\begin{thebibliography}{99}

\bibitem{Maldacena:1997re}
J.~M.~Maldacena, ``The large N limit of superconformal field
theories and supergravity,'' Adv.\ Theor.\ Math.\ Phys.\  {\bf 2},
231 (1998) [Int.\ J.\ Theor.\ Phys.\  {\bf 38}, 1113 (1999)]
[arXiv:hep-th/9711200]. S.~S.~Gubser, I.~R.~Klebanov and
A.~M.~Polyakov, ``Gauge theory correlators from non-critical
string theory,'' Phys.\ Lett.\ B {\bf 428}, 105 (1998)
[arXiv:hep-th/9802109].
E.~Witten, ``Anti-de Sitter space and holography,'' Adv.\ Theor.\
Math.\ Phys.\  {\bf 2}, 253 (1998) [arXiv:hep-th/9802150].

\bibitem{D'Hoker:2002aw}
E.~D'Hoker and D.~Z.~Freedman, ``Supersymmetric gauge theories and
the AdS/CFT correspondence,'' arXiv:hep-th/0201253.

\bibitem{Berenstein:2002jq}
D.~Berenstein, J.~Maldacena and H.~Nastase, ``Strings in flat
space and pp waves from N = 4 super Yang Mills,''
arXiv:hep-th/0202021.

\bibitem{Metsaev}
R.~R.~Metsaev,
``Type IIB Green-Schwarz superstring in plane wave Ramond-Ramond  background,''
Nucl.\ Phys.\ B {\bf 625}, 70 (2002)
[arXiv:hep-th/0112044].

\bibitem{Gubser:2002tv}
S.~S.~Gubser, I.~R.~Klebanov and A.~M.~Polyakov, ``A
semi-classical limit of the gauge/string correspondence,'' Nucl.\
Phys.\ B {\bf 636}, 99 (2002) [arXiv:hep-th/0204051].

\bibitem{Frolov:2002av}
S.~Frolov and A.~A.~Tseytlin, ``Semiclassical quantization of
rotating superstring in AdS(5) x S**5,'' JHEP {\bf 0206}, 007
(2002) [arXiv:hep-th/0204226].

\bibitem{Tseytlin:2002ny}
A.~A.~Tseytlin, ``Semiclassical quantization of superstrings:
AdS(5) x S**5 and beyond,'' arXiv:hep-th/0209116.

\bibitem{Russo:2002sr}
J.~G.~Russo,
``Anomalous dimensions in gauge theories from rotating strings in  AdS(5) x S(5),''
JHEP {\bf 0206}, 038 (2002) [arXiv:hep-th/0205244].

\bibitem{kruczenski}
M.~Kruczenski,
``A note on twist two operators in N = 4 SYM and Wilson loops in  Minkowski signature,''
JHEP {\bf 0212}, 024 (2002)
[arXiv:hep-th/0210115].

\bibitem{Gross:ju}
D.~J.~Gross and F.~Wilczek, ``Asymptotically Free Gauge Theories.
I,'' Phys.\ Rev.\ D {\bf 8} (1973) 3633.
D.~J.~Gross and F.~Wilczek, ``Asymptotically Free Gauge Theories.
2,'' Phys.\ Rev.\ D {\bf 9}, 980 (1974).

\bibitem{Iengo:2002tf}
R.~Iengo and J.~G.~Russo, ``The decay of massive closed
superstrings with maximum angular momentum,'' JHEP {\bf 0211}, 045
(2002) [arXiv:hep-th/0210245]. R.~Iengo and J.~G.~Russo,
``Semiclassical decay of strings with maximum angular momentum,''
JHEP {\bf 0303}, 030 (2003) [arXiv:hep-th/0301109].

\bibitem{Frolov:2003qc}
S.~Frolov and A.~A.~Tseytlin,
``Multi-spin string solutions in AdS(5) x S**5,''
Nucl.\ Phys.\ B {\bf 668}, 77 (2003)
[arXiv:hep-th/0304255].

\bibitem{tseytlin}
A.~A.~Tseytlin,
``On semiclassical approximation and spinning string vertex operators in  AdS(5) x S**5,''
Nucl.\ Phys.\ B {\bf 664}, 247 (2003)
[arXiv:hep-th/0304139].
S.~Frolov and A.~A.~Tseytlin,
``Quantizing three-spin string solution in AdS(5) x S**5,''
JHEP {\bf 0307}, 016 (2003)
[arXiv:hep-th/0306130].
S.~Frolov and A.~A.~Tseytlin,
``Rotating string solutions: AdS/CFT duality in non-supersymmetric  sectors,''
arXiv:hep-th/0306143.
G.~Arutyunov, S.~Frolov, J.~Russo and A.~A.~Tseytlin,
``Spinning strings in AdS(5) x S**5 and integrable systems,''
arXiv:hep-th/0307191.
N.~Beisert, S.~Frolov, M.~Staudacher and A.~A.~Tseytlin,
``Precision spectroscopy of AdS/CFT,''
arXiv:hep-th/0308117.

\bibitem{Armoni:2002xp}
A.~Armoni, J.~L.~Barbon and A.~C.~Petkou,
``Orbiting strings in AdS black holes and N = 4 SYM at finite  temperature,''
JHEP {\bf 0206}, 058 (2002)
[arXiv:hep-th/0205280].
\bibitem{Armoni:2002fr}
A.~Armoni, J.~L.~Barbon and A.~C.~Petkou,
``Rotating strings in confining AdS/CFT backgrounds,''
JHEP {\bf 0210}, 069 (2002)
[arXiv:hep-th/0209224].

\bibitem{Minahan:2002rc}
J.~A.~Minahan,
``Circular semiclassical string solutions on AdS(5) x S**5,''
Nucl.\ Phys.\ B {\bf 648}, 203 (2003)
[arXiv:hep-th/0209047].
N.~Beisert, J.~A.~Minahan, M.~Staudacher and K.~Zarembo,
``Stringing spins and spinning strings,''
arXiv:hep-th/0306139.

\bibitem{Rashkov:2002zt}
R.~C.~Rashkov and K.~S.~Viswanathan,
``Rotating strings with B-field,''
arXiv:hep-th/0211197.

\bibitem{alexandrova}
D.~Aleksandrova and P.~Bozhilov,
``On the classical string solutions and string / field theory duality,''
arXiv:hep-th/0307113.
D.~Aleksandrova and P.~Bozhilov,
``On the classical string solutions and string / field theory duality. II,''
arXiv:hep-th/0308087.

\bibitem{Ouyang:2002vg}
P.~Ouyang,
``Semiclassical quantization of giant gravitons,''
arXiv:hep-th/0212228.

\bibitem{Buchel:2002yq}
A.~Buchel, ``Gauge / string correspondence in curved space,''
arXiv:hep-th/0211141.

\bibitem{Pons:2003ci}
J.~M.~Pons and P.~Talavera, ``Semi-classical string solutions for
N = 1 SYM,'' arXiv:hep-th/0301178.

\bibitem{Alihahisha:2002fi}
M.~Alishahiha and A.~E.~Mosaffa,
``Circular semiclassical string solutions on confining AdS/CFT backgrounds,''
JHEP {\bf 0210}, 060 (2002)
[arXiv:hep-th/0210122].

\bibitem{Mateos:2003de}
D.~Mateos, T.~Mateos and P.~K.~Townsend,
``Supersymmetry of Tensionless Rotating Strings in $AdS_5 x S^5$, and Nearly-BPS Operators,''
arXiv:hep-th/0309114.

\bibitem{Alishahiha:2002sy}
M.~Alishahiha and M.~Ghasemkhani,
``Orbiting membranes in M-theory on AdS(7) x S(4) background,''
JHEP {\bf 0208}, 046 (2002)
[arXiv:hep-th/0206237].

\bibitem{Hartnoll:2002th}
S.~A.~Hartnoll and C.~Nunez,
``Rotating membranes on G(2) manifolds, logarithmic anomalous dimensions  and N = 1 duality,''
JHEP {\bf 0302}, 049 (2003)
[arXiv:hep-th/0210218].

\bibitem{ryang}
S.~Ryang,
``Rotating and orbiting strings in the near-horizon brane backgrounds,''
JHEP {\bf 0304}, 045 (2003)
[arXiv:hep-th/0303237].

\bibitem{Kahru}
S.~Kachru and E.~Silverstein,
``4d conformal theories and strings on orbifolds,''
Phys.\ Rev.\ Lett.\  {\bf 80}, 4855 (1998)
[arXiv:hep-th/9802183].

\bibitem{Klebanov:1998hh}
I.~R.~Klebanov and E.~Witten, ``Superconformal field theory on
threebranes at a Calabi-Yau  singularity,'' Nucl.\ Phys.\ B {\bf
536}, 199 (1998) [arXiv:hep-th/9807080].

\bibitem{Gubser:1998fp}
S.~S.~Gubser and I.~R.~Klebanov, ``Baryons and domain walls in an
N = 1 superconformal gauge theory,'' Phys.\ Rev.\ D {\bf 58},
125025 (1998) [arXiv:hep-th/9808075].

\bibitem{Klebanov:1999rd}
I.~R.~Klebanov and N.~A.~Nekrasov, ``Gravity duals of fractional
branes and logarithmic RG flow,'' Nucl.\ Phys.\ B {\bf 574}, 263
(2000) [arXiv:hep-th/9911096].

\bibitem{Klebanov:2000nc}
I.~R.~Klebanov and A.~A.~Tseytlin, ``Gravity duals of
supersymmetric SU(N) x SU(N+M) gauge theories,'' Nucl.\ Phys.\ B
{\bf 578}, 123 (2000) [arXiv:hep-th/0002159].

\bibitem{Klebanov:2000hb}
I.~R.~Klebanov and M.~J.~Strassler, ``Supergravity and a confining
gauge theory: Duality cascades and  chiSB-resolution of naked
singularities,'' JHEP {\bf 0008}, 052 (2000)
[arXiv:hep-th/0007191].

\bibitem{buchelS3}
A.~Buchel and A.~A.~Tseytlin,
``Curved space resolution of singularity of fractional D3-branes on  conifold,''
Phys.\ Rev.\ D {\bf 65}, 085019 (2002)
[arXiv:hep-th/0111017].

\bibitem{Sakai:2003wu}
T.~Sakai and J.~Sonnenschein,
``Probing flavored mesons of confining gauge theories by supergravity,''
arXiv:hep-th/0305049.

\bibitem{kruczenski2}
M.~Kruczenski, D.~Mateos, R.~C.~Myers and D.~J.~Winters,
``Meson spectroscopy in AdS/CFT with flavour,''
JHEP {\bf 0307}, 049 (2003)
[arXiv:hep-th/0304032].

\bibitem{Bershadsky:1995sp}
M.~Bershadsky, C.~Vafa and V.~Sadov, ``D-Strings on D-Manifolds,''
Nucl.\ Phys.\ B {\bf 463}, 398 (1996) [arXiv:hep-th/9510225].

\bibitem{Dasgupta:1998su}
K.~Dasgupta and S.~Mukhi, ``Brane constructions, conifolds and
M-theory,'' Nucl.\ Phys.\ B {\bf 551}, 204 (1999)
[arXiv:hep-th/9811139].

\bibitem{Dasgupta:1999wx}
K.~Dasgupta and S.~Mukhi, ``Brane constructions, fractional branes
and anti-de Sitter domain walls,'' JHEP {\bf 9907}, 008 (1999)
[arXiv:hep-th/9904131].

\bibitem{ohta2}
Kazutoshi Ohta, private communication.

\bibitem{uranga}
A.~M.~Uranga,
``Brane configurations for branes at conifolds,''
JHEP {\bf 9901}, 022 (1999)
[arXiv:hep-th/9811004].

\bibitem{ohta}
K.~Ohta and T.~Yokono,
``Deformation of conifold and intersecting branes,''
JHEP {\bf 0002}, 023 (2000)
[arXiv:hep-th/9912266].

\bibitem{Oh:2001bf}
K.~h.~Oh and R.~Tatar,
``Duality and confinement in N = 1 supersymmetric theories from geometric  transitions,''
Adv.\ Theor.\ Math.\ Phys.\  {\bf 6}, 141 (2003)
[arXiv:hep-th/0112040].

\bibitem{Hanany:1996ie}
A.~Hanany and E.~Witten,
``Type IIB superstrings, BPS monopoles,
and three-dimensional gauge  dynamics,'' Nucl.\ Phys.\ B {\bf
492}, 152 (1997) [arXiv:hep-th/9611230].

\bibitem{Bergshoeff:1995as}
E.~Bergshoeff, C.~M.~Hull and T.~Ortin, ``Duality in the type II
superstring effective action,'' Nucl.\ Phys.\ B {\bf 451}, 547
(1995) [arXiv:hep-th/9504081].

\bibitem{Tseytlin:1996bh}
A.~A.~Tseytlin,
``Harmonic superpositions of M-branes,'' Nucl.\
Phys.\ B {\bf 475}, 149 (1996) [arXiv:hep-th/9604035].

\bibitem{Tseytlin:1997cs}
A.~A.~Tseytlin,
``Composite BPS configurations of p-branes in 10
and 11 dimensions,'' Class.\ Quant.\ Grav.\  {\bf 14}, 2085 (1997)
[arXiv:hep-th/9702163].

\bibitem{Gauntlett:1997cv}
J.~P.~Gauntlett,
``Intersecting branes,'' arXiv:hep-th/9705011.

\bibitem{Andreas:1998hh}
B.~Andreas, G.~Curio and D.~Lust,
``The Neveu-Schwarz five-brane
and its dual geometries,'' JHEP {\bf 9810}, 022 (1998)
[arXiv:hep-th/9807008].

\bibitem{Ohta03}
N.~Ohta and T.~Shimizu, ``Non-Extreme Black Holes From
Intersecting M-Branes,'' Int. J. of Mod. Phys. {\bf A13} (1998)
1305, [hep-th/9701095]; N.~Ohta, ``Intersection Rules for
Non-Extreme $p$-Branes,'' Phys. Lett. {\bf B403} (1997) 218 --
224, [hep-th/9702164]; N.~Ohta and Jian-Ge Zhou, ``Towards the
Classification of Non-Marginal Bound States of M-Branes and Their
Construction Rules,'' Int. J. of Mod. Phys. {\bf A13} (1998) 2013,
[hep-th/9706153],

\bibitem{youm}
D.~Youm,
``Localized intersecting BPS branes,''
arXiv:hep-th/9902208.

\bibitem{hashimoto}
A.~Hashimoto,
``Supergravity solutions for localized intersections of branes,''
JHEP {\bf 9901}, 018 (1999)
[arXiv:hep-th/9812159].

\bibitem{Korchemsky:1988si}
G.~P.~Korchemsky,
``Asymptotics Of The Altarelli-Parisi-Lipatov Evolution Kernels Of Parton Distributions,''
Mod.\ Phys.\ Lett.\ A {\bf 4}, 1257 (1989).

\bibitem{Korchemsky:1992xv}
G.~P.~Korchemsky and G.~Marchesini,
``Structure function for large x and renormalization of Wilson loop,''
Nucl.\ Phys.\ B {\bf 406}, 225 (1993)
[arXiv:hep-ph/9210281].

\bibitem{Gursoy:2002tx}
U.~Gursoy, C.~Nunez and M.~Schvellinger,
``RG flows from Spin(7), CY 4-fold and HK manifolds to AdS, Penrose  limits and pp waves,''
JHEP {\bf 0206}, 015 (2002)
[arXiv:hep-th/0203124].

\bibitem{Schwarz:qr}
J.~H.~Schwarz, ``Covariant Field Equations Of Chiral N=2 D = 10
Supergravity,'' Nucl.\ Phys.\ B {\bf 226}, 269 (1983).

\bibitem{Tran:jj}
T.~A.~Tran,
``Gauged Supergravities From Spherical Reductions,''
UMI-30-33892

\end{thebibliography}
\end{document}